\documentclass[paper]{elsart}
\usepackage{amssymb,latexsym,amsmath}
\usepackage{graphicx, epsf}

\newcommand{\be}{\begin{equation}}
\newcommand{\ee}{\end{equation}}
\newcommand{\bea}{\begin{eqnarray}}
\newcommand{\eea}{\end{eqnarray}}
\newcommand{\beas}{\begin{eqnarray*}}
\newcommand{\eeas}{\end{eqnarray*}}

\newcommand{\pardt}{\partial_t}
\newcommand{\pardx}{\partial_x}
\newcommand{\pardxx}{\partial_{xx}}
\newcommand{\pardy}{\partial_y}
\newcommand{\pardY}{\partial_Y}
\newcommand{\pardYY}{\partial_{YY}}

\renewcommand {\theequation}{\thesection.\arabic{equation}}
\renewcommand {\thetable}{\thesection.\arabic{table}}

\begin{document}

\begin{frontmatter}

\title{Singularity formation for Prandtl's equations}
\author[authorlabel]{M. Sammartino\corauthref{marco}},
\ead{marco@math.unipa.it}
\author[authorlabel]{V. Sciacca},
\ead{sciacca@math.unipa.it}
\author[authorlabel]{F. Gargano}
\ead{gargano@math.unipa.it}

\maketitle
\date{}


\address[authorlabel]{University of Palermo, Department of Mathematics\\  Via Archirafi 34, 90123 Palermo, Italy.}%
\corauth[marco]{Corresponding author.}

\begin{abstract}
We consider Prandtl's equations for the impulsively started disk and
follow the process of the formation of the singularity  in the
complex plane using the singularity tracking method.
We classify Van Dommelen and Shen's singularity  as a cubic root
singularity.

We introduce a class of initial data which  have
a dipole singularity in the complex plane.
These data  are uniformly bounded in  $H^1$ and lead to an earlier singularity formation.
The blow up time, which seems to be short, behaves as the
blow up time of Burger's equation with data in the same class.

The presence of a small viscosity in the streamwise
direction changes the behavior of the singularities.
They stabilize at a distance from the real axis which depends on the amount of the
viscosity.
\end{abstract}

\begin{keyword}
Prandtl's equations \sep Separation \sep Spectral Methods \sep
Complex singularities \sep  Blow--up time \sep Regularizing viscosity
\MSC  76D10 \sep 35Q35 \sep 35A20 \sep 65M70

\end{keyword}

\end{frontmatter}

\section{Introduction}
\setcounter{equation}{0}

The aim of this paper is to perform a numerical study of the process
of singularity formation for the solutions of Prandtl's equations.
The main reason that motivates this study is that the
mathematical theory of Prandtl's equations is still unclear, as
no well posedness result, under general hypotheses on the data,
is available.
It would be an advance in the mathematical theory of fluid dynamics
to prove that the solutions of Prandtl's equations remain (for a short time) regular
when the initial and boundary data have  (for example)
a Sobolev type of regularity; or to disprove this, constructing a class of data,
bounded in a Sobolev space, for which  a zero time blow-up of the solutions occurs.

The main result of this paper is a numerical evidence of the
ill-posedness of Prandtl's equations in $H^1$.
Other results we shall obtain with our investigation are the following.
First  the blow-up of Prandtl's solution for the impulsively started
disk will be shown to be the consequence of a complex singularity, of
cubic-root type, hitting the real $x$ axis; this result  had already
been found in \cite{DLSS06}, but in this paper we give further
evidences also using more refined numerical schemes.
Second we shall see that this structure of the singularity is robust
with respect to perturbations (belonging to a class we shall specify later)
of the initial datum as well of the matching datum.
Third the blow-up time for the solution will be shown to be related to the
$\sup$ of the derivative of the initial datum, at least when the complex
singularity is relatively close to the real axis.
Finally the presence of a small viscosity in the streamwise direction
is able to prevent the formation of the singularity (a result already
present in \cite{DLSS06}), while the number of Fourier modes
effectively excited tends to be linearly dependent on the streamwise
viscosity.

In the rest of this Introduction we shall briefly recall what the Van Dommelen
and Shen  singularity is, and give a short account
on the mathematical theory of Prandtl's equations.
For more details on these topics we shall refer the reader to the review papers
available in the literature.

\subsection{Separation and singularity in the boundary layer}

Boundary layer separation is the most striking phenomenon that one encounters
considering the interaction of a slightly viscous (or, said differently,
with high Reynolds number) flow with a solid boundary.
Unsteady separation  has been puzzling the researchers for many decades.
The importance of this phenomenon stays in the fact that separation  is the mechanism
through which the vorticity generated at the boundary is ejected into the
outer flow  and it is therefore considered  as one of the responsible
of the transition to turbulence in wall bounded fluids.

In this contest the case of the 2D flow past an impulsively started
disk has attracted great interest because it served as a case study
where (since the early stages of the investigations) the appearance
of reverse flow, recirculation, separation, vorticity shedding were
proposed as a possible universal route to the onset of turbulence.
The seminal work of Van Dommelen and Shen  \cite{VDS80,VDS82} (see
also e.g. \cite{VDC89}) clarified how separation is linked
to the appearance of a singularity in the solution of Prandtl's
equations.
In their Lagrangian framework the singularity appeared as
a fluid particle being squashed in the streamwise direction to zero
thickness, therefore causing (in absence of normal pressure
gradients, as it is the case for Prandtl's equations) an eruption in
the normal direction.
Subsequent calculations by different authors
both in Lagrangian coordinates as well using finite difference or
spectral \cite{Ing84} methods in Eulerian coordinates confirmed the
results of Van Dommelen and Shen, and the fact that Prandtl's
equations develop singularity became an accepted statement in the
literature.

More problematic is the relationship between the process
of separation for the high Reynolds number Navier-Stokes flow  and the
singularity of Prandtl's solution.
Although the qualitative picture of the Navier-Stokes solutions, with the
appearance of high gradients in the streamwise velocity, resembles the
structure of the singularity in  Prandtl's solution, a more careful
quantitative study reveals that the interaction of the boundary layer
flow with the outer flow sets in well before the singularity time for Prandtl's
equations (see \cite{Cas00}),
at least for the Reynolds numbers that have been tested so far, which are
about $10^7$.

It is almost impossible here to review the relevant literature regarding
the development of the boundary layer theory after Van Dommelen
and Shen's breakthrough, and we refer the interested reader to
\cite{Cow01}.

\subsection{The mathematical theory of Prandtl's equations}

The fact that the appearance of a singularity seems related to a
physical phenomenon (separation) that experimentalists have
observed for decades makes the mathematical theory of Prandtl's
equations relevant also from the applied point of view.
Moreover the well- or the ill-posedness of Prandtl's equations
is related to a fundamental question of theoretical fluid
dynamics, i.e. the convergence of the Navier--Stokes solutions to
the solutions of the Euler equations away from boundaries.
The first results on the existence and the uniqueness issue for the
solutions of Prandtl's  equations were obtained by Oleinik and her
coworkers (see e.g. \cite{O66}).
Basically Oleinik's theory (which is reviewed in great detail in
\cite{OS99}) deals with monotonous data, in the sense that
e.g. $\pardy u_0>0$, $u_0$ being the initial datum.
In this case Oleinik and her coworkers were able to get short time
existence of regular solutions (or long time existence for small
domains).
Recently, adding the hypothesis of a favorable pressure
gradient (i.e. $\pardx p\leq 0$), Xin and Zhang in \cite{XZ04}
proved the long time existence of weak solutions.
Other  results, concern the case of analytic data.
In \cite{SC98a} (see also \cite{Asano88b}), the authors proved
the short time existence and uniqueness when the data are
analytic; this result was improved in \cite{CLS01,LCS03} where it is
required analyticity in the streamwise direction while, using the
regularizing effect of the viscosity,  only $C^2$ regularity in
the normal direction is found to be necessary.
On the other side there is the result of E and Engquist \cite{EE97}
which  for a flat boundary proved that, if the initial data is of the form
$u_0=-xb_0(x,y)$, and if $\pardx u_0(x=0,y)$ satisfies a technical
assumption, then at $x=0$ the solution terminates in a blow up of
the tangential derivative.
For a review on the mathematical theory of Prandtl's equations,
see \cite{CS00,E00} and also \cite{BT07}.

Regarding the related problem of the convergence of Navier--Stokes solutions in the
zero viscosity limit, we mention the result in \cite{SC98b} where  for analytic
data the authors prove  convergence  to Euler and Prandtl's solutions (away and close to the
boundary respectively), and the criteria of Kato \cite{Ka84} and of Temam and Wang \cite{TW97}.
Assuming an {\em a priori} estimate on the solution close to the boundary,
(on the energy dissipation in \cite{Ka84} or on the pressure
gradient in \cite{TW97}), one gets convergence to the Euler solution
away from the boundary.

On the other hand Grenier, in \cite{Gren00a}, proved that there are initial profiles
(those profiles for which Euler equations are linearly unstable)
for which the exponential growth of modes of size $\mbox{Re}^{1/2}$
(i.e. a phenomenon linked to Rayleigh instability) does not allow the
solution to have the form of a matched asymptotic expansion between a Prandtl
solution and an Euler solution.
The findings of Brinckman and Walker are probably somehow related to
this result. In  \cite{BW01}, using the full 3D Navier--Stokes
equations, they studied the response of the boundary layer to an
array of counter-rotating vortices placed in the outer flow.
Their goal was to mimic the mechanism through which vorticity regenerates
in the boundary layer. They found that the time at which the
Rayleigh instability caused rapidly growing oscillations inside the
boundary layer was shorter than the time at which  Van Dommelen
and Shen's singularity developed which posed again the question
whether (or up to what time) the Navier--Stokes solutions are
correctly described by  Prandtl's equations.
Although more recent (at higher resolution) calculations \cite{OC05} seem to cast some
doubt on the prediction of \cite{BW01} of an early development of
instabilities, Brinckman and Walker {\em  pose an important
problem}, see the discussion in \cite{Cow01} where the possibility
of the Rayleigh instabilities winning the race with Van Dommelen and
Shen's singularity is suggested also for data with exponentially
decaying spectra (i.e. analytic data).

\subsection{Plan of the paper}

In this paper we  investigate, mostly in the case study of
the $2D$ impulsively started disk, the process of the
singularity formation for Prandtl's equations using the
singularity tracking technique.
This methodology was initiated in \cite{SSF83}, and is based on the
idea that the blow-up of the solution (or of its derivative) of a PDE
is preceded by the appearance, in the plane of the complexified
spatial variable, of a complex singularity.
The real blow-up is the consequence of the singularity hitting the real axis.
The location of the complex variable can be determined studying the
asymptotic behavior of the Fourier spectrum.
Although the method was originally introduced to investigate the
possibility of the finite time blow up of the $3D$ Euler solutions
(see e.g. \cite{Caf93,FMB03,MBF05,PMFB06}), it turned out to be a powerful
tool to characterize the process of singularity formation for many
PDEs, see for example \cite{CBT99,GPS98,PS98,Sh92}.

In  Section \ref{NM} we describe all the various numerical
method we used to solve Prandtl's equations in the case of  the 2D
impulsively started disk.
They are all spectral method in the streamwise  direction
while in the normal direction we have used both finite-differences
and spectral methods (using Chebychev polynomials).
For the discretization in time  various methods have been used.
All these numerical procedures lead to the same conclusions,
and this gives some reliability to our results.
In this Section we shall also describe the singularity tracking method,
both in $1D$ and in $2D$.

In Section \ref{VDSsing} we study the Van Dommelen and Shen singularity.

In Section \ref{ESF} we propose to study Prandtl's equations imposing a class of
real analytic (with respect to the streamwise direction while it is $C^2$ w.r.t. the
normal variable) initial data that have an algebraic singularity placed at some distance
$\delta_0$ from the real axis.
To be more precise we take Van Dommelen and Shen's solution at time $t=1.5$
(when their singularity is quite far away from the real axis: we recall that
the VDS singularity time is $t\approx 3$) and add a singularity
placed at distance $\delta_0$ from the real axis.
The spectrum of our data behaves like $|k|^{-5/3}$.
This means that, however small $\delta_0$ is, all the data in this class have
$H^1$ norm bounded by some constant.
Moreover the singularity we are placing in the complex plane can be thought of as a couple
of singularities of opposite sign infinitely close and infinitely strong.
We call this singularity ``dipole singularity''.
In Section \ref{ESF} we show that this singularity travels toward the real axis to lead
to a blow up of the solution in a time shorter than the typical Van Dommelen and
Shen's singularity time.

In Section \ref{casestudy} we study how  Burger's equation develops singularity
for the dipole singularity initial data.
We see how, in the well understood case of  Burger's equation
(in fact one knows the relationship between the singularity time and the
$\sup$ norm of the derivative of the initial datum), one can make arbitrarily
short the singularity time picking the $\delta_0$ sufficiently small.

In Section \ref{Discussion} we see that the situation seems to be
the same for Prandtl's equation.

In Section \ref{regularized} we propose to study Prandtl's equations
with the presence of the regularizing viscosity in the streamwise direction.
We will see that a small viscosity in $x$ is able to prevent the singularity formation.
The singularity, still headed to the real axis, fails to reach it and stabilizes at
a distance that depends on the amount of viscosity in $x$.
In this section we also clarify how the class of initial data introduced in
Section \ref{ESF} has the physical meaning of a couple of counter--rotating
vortices placed inside the boundary layer.

\section{The equations and the numerical methods}\label{NM}
\setcounter{equation}{0} In this section first we shall introduce
Prandtl's equations in the case of an impulsively started disk
in a uniform background flow. Second we shall discuss the various
numerical methods we have used to solve the equations. Finally we
shall give a  brief review of the ideas behind the singularity
tracking method.

\subsection{The equations}
Prandtl's equations describe the behavior of a fluid close to a
physical boundary in the zero viscosity limit. These equations can
be obtained as a formal asymptotic limit of the Navier--Stokes
equations assuming that close to the boundary there is a rapid
adjustment between the outer inviscid flow and the inner viscous
flow.
If in the Navier-Stokes equations one rescales the variable normal
to the boundary as $Y=y/\sqrt{\nu}$, to the first order in $\sqrt{\nu}$
one gets the Prandtl equations.
In the case of a disk  impulsively started in a uniform background
flow $U=\sin{(x)}$, Prandtl's equations are:

\bea
\pardt u+u\pardx u +v\pardY u-U_{ }\pardx U_{ }&=&\pardYY u \qquad U=\sin{(x)}
\label{P1}\\
\pardx u +\pardY v &=&0  \label{P2}\\
u(x,0,t)=v(x,0,t)&=&0  \label{P3}\\
u(x,\infty,0)&=&U_{ }  \label{P4}\\
u(x,y,0)&=&U \, .\label{P5}
\eea

The above equations have to be solved for $(x,Y)\in [0,2\pi]\times[0,\infty[$
for the variable $u$.
In fact, the incompressibility condition \eqref{P2} allows to recover the rescaled
normal velocity $v$ through an integration in the normal variable.
Equation \eqref{P3} and Eq.\eqref{P4} give the boundary condition and the matching
condition with the outer flow, while \eqref{P5} is the initial condition.

\subsection{The numerical methods I: mixed spectral-finite difference}

We  solve the above equations in the computational domain $[0,
2\pi]\times[0,\overline{Y}]$.
One has to choose $\overline{Y}$ big
enough so that the solution $u(x,\overline{Y},t)$ has matched
exponentially the matching condition \eqref{P4} (i.e.
$u(x,\overline{Y},t)-U$ must be small). In our calculations we have
chosen $\overline{Y}=20$, as in \cite{Hong02,HoHu03} or
\cite{DLSS06}, which is sufficient to ensure the matching between
$u$ and $U$.

To solve the equations in [0,T] one can use a  spectral method for the variable $x$
and finite differences for the variable $Y$.
Namely, denoting with $\Delta Y=20/(M+1)$ and with $\Delta t=T/N$, we approximate the solution
as:
$$
u(x,j\Delta Y,n\Delta t)\approx \sum_{k=-K/2}^{K/2} u^n_{k,j}e^{ikx}
$$
The diffusion term $\pardYY u$ has been treated using the usual three points rule.
We have now to specify the way we have discretized the time
derivative $\pardt u$,  and the nonlinear terms
$\pardx F(u)$ and $\pardY G(u)$ where $F(u)\equiv u^2$ and $G(u)\equiv uv$.
Regarding the nonlinear term involving the $x$ derivative we have always used
the usual pseudo-spectral approximation involving multiplication in the physical space
via an inverse FFT, and using the usual $3/2$--rule to handle aliasing effects \cite{CHQZ06}.
As far as the other terms are concerned we have used several algorithms.
A possibility is to use the one step Euler method for the time derivative,
the Crank--Nicolson method for the diffusion term and to use
a Lax-Wendroff approximation for the nonlinear term involving the $Y$ derivative
in the form:
\be
\begin{array}{rcl}
[\pardY G(u)]^n_{k,j}&=& \lambda \left( [G(u)]^n_{k,j-1/2}-[G(u)]^n_{k,j+1/2}\right) \qquad
\lambda\equiv \Delta t/\Delta Y \, , \\
\left[G(u)\right]^n_{k,j+\frac{1}{2}}&=& \frac{1}{2} \left\{v^n_{j+\frac{1}{2}} * \left[ u^n_j +u^n_{j+1} -
\lambda v^n_{j+\frac{1}{2}} * \left( u^n_{j+1} -u^n_j\right) \right]\right\}_k  \, ,\\
v^n_{k,j+\frac{1}{2}}&=&\frac{1}{2}(v^n_{k,j}+v^n_{k,j+1})  \, ,
\end{array} \label{LW}
\ee
where by $*$ we have denoted the (pseudo)-spectral multiplication.
The scheme just described, which is the spectral (in $x$) version of the scheme used in
\cite{Hong02,HoHu03}, has been used in \cite{DLSS06} where the convergence properties are
described.
An improvement on the above scheme is obtained using the implicit--explicit midpoint
method $(1,2,2)$ (see \cite{ARS97} for the details) based on the following
pair of implicit (for the diffusion term) and explicit (for the nonlinear
terms) Runge--Kutta schemes:
\be
\begin{tabular}{l|ll}
0   & 0& 0 \\
1/2 & 0& 1/2 \\
\hline
    & 0 & 1
\end{tabular}\qquad \quad
\begin{tabular}{l|ll}
0   & 0& 0 \\
1/2 & 1/2& 0 \\
\hline
    & 0 & 1
\end{tabular}\label{122}
\ee

The convergence properties of this scheme at various times (well
before the singularity time, just before the singularity time and
at the singularity time) are given in Table \ref{convLW122}.
Similar results are obtained with the use of higher order IMEX schemes.
Notice that in the convergence tables we are reporting the $L^2$ and $L^1$ errors
have been computed up to the area
of the computational domain, i.e. as:
$$
\|f\|^2_{L^2} \equiv \sum_{k,j} |f_{k,j}|^2 \qquad
\|f\|_{L^1} \equiv \sum_{k,j} |f_{k,j}| \; .
$$
To get the real $L^2$ and $L^1$ norms one should consider the factors $A^{1/2}$ and $A$
with $A\approx 125.6$.

\begin{table}[ht]
\begin{tiny}
\begin{center}
\caption{Convergence of the numerical scheme based on the Lax--Wendroff approximation
of $\pardY G(u)$ and on the IMEX scheme \eqref{122}.
We have obtained similar results coupling the Lax--Wendroff approximation with the IMEX scheme
(2,3,3) given in  \eqref{233} and the IMEX scheme (3,4,3) described in \cite{ARS97} pp.158--159.}
\label{convLW122}
\vskip.1cm
\begin{tabular}{|c|c|c|c||c|c|c||c|c|c|}\hline
Grid   & \multicolumn{3}{|c||}{$T=1.54$} &\multicolumn{3}{|c||}{$T=2.85$} &\multicolumn{3}{|c|}{$T=2.97$}\\  \hline
($N\times K\times M$) & $L^2$    & $L^1$     & $L^\infty$
& $L^2$    & $L^1$     & $L^\infty$
& $L^2$    & $L^1$     & $L^\infty$
\\ \hline
$10^4\times 32\times 127$         & $2.5\cdot 10^{-4}$ & $1.5\cdot 10^{-4}$   & 0.001
                                  & 0.06               & 0.04                 & 0.207
                                  & 0.09                & 0.06                 & 0.345
\\ \hline
$2\cdot 10^4\times 64\times 255$  & $6.3\cdot 10^{-5}$   & $3.7\cdot 10^{-5}$ & $2.5\cdot 10^{-4}$
                                  & 0.029               & 0.018               & 0.158
                                  & 0.057               & 0.039               & 0.347
 \\ \hline
$4\cdot 10^4\times 128\times 511$ & $1.6\cdot 10^{-5}$ & $9.5\cdot 10^{-6}$ & $6.6\cdot 10^{-5}$
                                  & 0.012               & 0.006              & 0.113
                                  & 0.037               & 0.023               & 0.353
 \\ \hline
\end{tabular}
\end{center}
\end{tiny}
\end{table}
Instead of the finite--differences approximation
\eqref{LW}, we have also used a spectral method to calculate $v\pardY u$.
In fact we extended $u$ to $Y\in[20,40]$ to make it periodic, passed to the Fourier
space in $Y$ and evaluated the derivative $\pardY u$, passed back to the physical space
to finally multiply times $v$ in the physical space in $Y$.
Due to the initial discontinuity  of the initial datum (notice the incompatibility
between the boundary condition \eqref{P3} and the initial condition \eqref{P5}) this method
works only after the viscosity has made the solution smooth in $Y$.
To handle this problem first we have  resolved the initial layer (let us say up to time
$t=0.05$) using one of the previous methods based on finite--differences at high
resolution and thereafter used the spectral approximation of $v\pardY u$.
Coupling this method with the IMEX scheme \eqref{122} we get the convergence
Table \ref{convSpec122}.

Similar convergence properties are obtained if one couples the spectral evaluation of
$v\pardY U$ with the IMEX scheme $(2,3,3)$ \cite{ARS97} given by the
pair of two stages implicit and three stages explicit  Runge--Kutta schemes:
\be
\begin{tabular}{l|ll}
$\gamma$   & $\gamma$   & 0      \\
$1-\gamma$ &$1-2\gamma$ & $\gamma$ \\
\hline
         &    1/2   & 1/2
\end{tabular}\qquad \quad
\begin{tabular}{l|lll}
0        & 0            & 0           & 0 \\
$\gamma$   & $\gamma$   & 0           & 0 \\
$1-\gamma$ & $\gamma-1$ &$2(1-\gamma)$& 0 \\
\hline
         & 0            &$1/2$        &$1/2$
\end{tabular}\label{233}
\ee
with $\gamma=(3+\sqrt{3})/6$.

\begin{table}[ht]
\label{convSpec122}
\caption{Convergence of the numerical scheme based on the spectral approximation
of $v\pardY u$ and on the Runge--Kutta IMEX scheme \eqref{122}.
Similar result are obtained using the Runge--Kutta IMEX scheme \eqref{233} and
the scheme (3,4,3) given in \cite{ARS97} pp.158--159. }
\begin{tiny}
\begin{tabular}{|c|c|c|c||c|c|c||c|c|c|}\hline
Grid   & \multicolumn{3}{|c||}{$T=1.5$} &\multicolumn{3}{|c||}{$T=2.85$} &\multicolumn{3}{|c|}{$T=2.97$}\\  \hline
$N\times K\times M$            & $L^2$         & $L^1$     & $L^\infty$
& $L^2$        & $L^1$         & $L^\infty$
& $L^2$        & $L^1$         & $L^\infty$
\\ \hline
$10^4\times 32\times 127$       &$1.5\cdot 10^{-4}$&$9.04\cdot  10^{-5}$ &$9.3\cdot  10^{-4}$
                                & 0.04             & 0.019               & 0.185
                                & 0.06             & 0.03                & 0.232
\\ \hline
$2\cdot10^4\times 64\times 255$ &$3.3\cdot 10^{-5}$&$1.65\cdot
10^{-5 }$  & $2.2\cdot 10^{-4}$
                                & 0.02             &0.009                 & 0.181
                                & 0.04             & 0.02                 & 0.326
 \\ \hline
$4\cdot10^4\times 128\times 511$&$8.0\cdot 10^{-6}$&$3.4\cdot 10^{-6}$   & $5.3\cdot 10^{-5}$
                                &0.008             &0.003                & 0.13
                                & 0.027            & 0.01                & 0.35
 \\ \hline
\end{tabular}
\end{tiny}
\end{table}
We have also used the Runge--Kutta scheme (3,4,3) given in \cite{ARS97} obtaining
similar results. This scheme has the advantage of allowing to take a larger
time step size, due to its stability properties.

All these methods give consistent results with
each other.
In Fig. \ref{VDSfig} we report the solution of
Prandtl's equation with initial datum given by \eqref{P5} computed
with the IMEX scheme $(1,2,2)$ of \eqref{122}, with $K=1024$ and
$M=1023$. All the computations are in agreement with  Van Dommelen and
Shen's (VDS) results \cite{VDS80}.


\subsection{The numerical methods II: fully  spectral}

We consider the same computational domain $[0,2\pi]\times[0, \overline{Y}]$
with $ \overline{Y}=20$ but we use the  Chebychev spectral approximation in the normal variable
$Y$.
To accomplish this, first we use the linear change of variable $Y\rightarrow w$ given by
$Y=(w\overline{Y}+\overline{Y})/2$ so that the computational domain becomes $[0,2\pi]\times[-1,1]$.
Second we use the fully spectral approximation:
\be
u(x,w,t)\thickapprox \sum\limits_{k=-K/2}^{k=K/2}\sum\limits_{j=0}^{j=M}u_{kj}(t)e^{ikx}T_j(w)\; ,
\label{ChebT}
\ee
where $T_j$ are the Chebychev polynomials of the first kind.
Finally, introducing the variable $\zeta$ defined as $w=\cos{(\zeta)}$, we write the above expression
as:
\be
u(x,\zeta,t)\thickapprox \sum\limits_{k=-K/2}^{k=K/2}\sum\limits_{j=0}^{j=M}u_{kj}(t)e^{ikx} \cos{(j\zeta)}\; ,
\label{ChebCos}
\ee
which, inserted in Prandtl's equation, through the usual $\tau$--method \cite{Boyd01,CHQZ06},
gives a fully spectral approximation scheme.
The boundary condition are implemented through the equations:
\begin{equation}
\sum\limits_{j=0}^{M}(-1)^j u_{kj}=0, \qquad \sum\limits_{j=0}^{M} u_{kj}=U_{k}, \qquad k=-K/2 \dots K/2,
\end{equation}
where $U_k$ are the Fourier components of the matching datum $U(x)$ defined in \eqref{P1}.

The non-linear terms $F(u)=u^2$ and $G(u)=uv$
are calculated by the usual pseudo-spectral approximation and  the aliasing effects
are handled by the  $3/2$--rule.

Discretization in time is done treating explicitly the convective terms and treating diffusion
by the Crank-Nicolson approximation.
In particular we have used a semi-implicit Runge-Kutta scheme, the two stage
RK2/CN, see \cite{Boyd01}.
We have also tried the three-stage semi-implicit Runge-Kutta scheme RK3/CN and the semi-implicit
Adams-Bashforth schemes of second and
third order but, as explained in \cite{Boyd01}, we have noticed that the  second-order RK2/CN
is more stable as it allows to take bigger time steps.
In fact we have imposed the CFL condition as $\Delta t=\frac{5}{2 M \max|v|}$.
In Table \ref{RK2/CN} we show the convergence properties of this scheme.

\begin{table}[ht]
\label{RK2/CN}
\caption{Convergence of the numerical scheme based on the full spectral approximation
and on the RK2/CN scheme.}
\begin{tiny}
\begin{tabular}{|c|c|c|c||c|c|c||c|c|c|}\hline
Grid   & \multicolumn{3}{|c||}{$T=1.5$} &\multicolumn{3}{|c||}{$T=2.85$} &\multicolumn{3}{|c|}{$T=2.97$}\\  \hline
$N\times K\times M$            & $L^1$         & $L^2$     & $L^\infty$
& $L^1$        & $L^2$         & $L^\infty$
& $L^1$        & $L^2$         & $L^\infty$
\\ \hline
$10^4\times 32\times 128$       &$1.59\cdot 10^{-4}$&$5.96\cdot  10^{-5}$ &$4.3\cdot  10^{-5}$
                                & 0.085             & 0.028               & 0.051
                                & 0.686             & 0.255                & 0.227
\\ \hline
$2\cdot10^4\times 64\times 256$ &$8.29\cdot 10^{-5}$&$2.98\cdot
10^{-5 }$  & $2.18\cdot 10^{-5}$
                                & 0.039             &0.013                 & 0.026
                                & 0.331             &0.132                 & 0.177
 \\ \hline
$4\cdot10^4\times 128\times 512$&$4.2\cdot 10^{-5}$&$1.84\cdot 10^{-5}$   & $1.38\cdot 10^{-5}$
                                &0.019             &0.006               & 0.012
                                &0.145             &0.052               & 0.087
 \\ \hline
\end{tabular}
\end{tiny}
\end{table}

\subsection{Singularity tracking}
If an analytic function $u(z)$ has an algebraic singularity at the
complex location $z^*=x^*+i\delta$, and if at the singularity the behavior is
$u(z)\approx (z-z^*)^\alpha$, then the Fourier spectrum of $u(z)$
has the following asymptotic behavior \cite{CKP66}:
\be
u_k\sim
|k|^{-(1+\alpha)}\exp{(- \delta |k|)}\exp{(ix^*k)}
\label{asyspectrum}
\ee
Therefore if one estimates the rate
$\delta$ of the exponential decay of the spectrum one gets the
distance of the complex singularity from the real axis;
the  estimate of the period of the oscillations of the spectrum  gives
the real location $x^*$ of the singularity.
Resolving the rate of
algebraic decay $1+\alpha$, one can classify the singularity type.
In particular, when one deals with a PDE with solution $u(t)$,
one can detect, with a good degree of reliability, the time at which the
singularity develops as given by
the first $t_c$ at which $\delta(t_c)=0$.

It is possible to extend the above technique to bi-variate functions, and we refer to \cite{MBF05,PMFB06}
for more details.
Let $u(x_1,x_2)$ a function with Fourier expansion:
\begin{equation}
u(x_1,x_2)=\sum_{k_1,k_2} u_{k_1 k_2} e^{i k_1 x_1} e^{i k_2 x_2} \; . \label{bi_fourier}
\nonumber
\end{equation}
If in the Fourier $\mathbf{k}$--space one considers those modes $(k_1, k_2)$ such that
$k_1= k \cos{(\theta)}$,  $k_2= k \sin{(\theta)}$ where $k=|(k_1,k_2)|$,
then the Fourier coefficients, when $k\rightarrow\infty$, have the following asymptotic  behavior:
\begin{equation}
u_{k_1 k_2}\approx  k^{-\left(\alpha(\theta)+1\right)} e^{-\delta(\theta) k} e^{i k x^*(\theta)} \qquad
(k_1, k_2)= k( \cos{(\theta)},  \sin{(\theta)}) \; .
 \label{Fourier_asymp_theta}
\end{equation}
The width of the analyticity strip $\overline{\delta}$ is  the minimum over all directions $\theta$, i.e.
$\delta^*=\min_{\theta}\delta(\theta)$.

A second way to extend the singularity tracking method to bi-variate functions is to define
the shell-summed Fourier amplitudes:
$$
A_K \equiv \sum_{K\leq |(k_1,k_2)| < K+1} \left|u_{k_1 k_2} \right|\; , 
$$
which is a kind of discrete angle average of the Fourier coefficients.
The asymptotic behavior of these amplitudes is:
$$
A_K\approx CK^{-\left(\alpha_{Sh}+1/2\right)} \exp{\left(-\delta_{Sh} K\right)} \qquad \mbox{when}\quad
K\rightarrow\infty \; ,
$$
where $\delta_{Sh}$ gives the width of the analyticity strip while the algebraic
prefactor $\alpha_{Sh}$ gives informations on the nature
of the singularity.

As pointed out in \cite{PMFB06}, using a steepest descent argument, one can see that
the two techniques are equivalent.
In fact, if one denotes with $\theta^*$ the angle where  $\delta(\theta)$
takes its minimum (i.e. $\delta^*=\delta(\theta^*)$),
one has that $\delta_{Sh}=\delta(\theta^*)$ and that $\alpha_{Sh}=\alpha(\theta^*)-1/2$.

An interesting situation is when the most singular direction
coincides with one of the coordinate axes, e.g. $\theta^*=0$. In
this case it is easy to see that, to evaluate the width of the strip
of analyticity, one can consider the variable  $x_2$ as a
parameter and adopt the following procedure.
First take the Fourier expansion relative to the variable $x_1$:
$$
u(x_1,x_2)=\sum_{k_1} u_{k_1}(x_2) e^{-ik_1x_1} \, ;
$$
second use that the spectrum has the asymptotic behavior:
$$
u_{k_1}(x_2) \approx k_1^{-(\alpha(x_2)+1)} e^{-\tilde{\delta}(x_2)k_1} \;
$$
finally get the width of analyticity as the minimum of $\tilde{\delta}(x_2)$,
i.e.: 
$$\delta_{Sh}= \min_{x_2}\tilde{\delta}(x_2),$$ 
and using again steepest descent argument, one can see that 
$$\alpha_{Sh}= \min_{x_2}\tilde{\alpha}(x_2),$$ 
when $\theta^*=0$.

The above procedures needs high numerical
precision and in fact, in the calculations we shall present,
we have used a 32--digits precision (using the
ARPREC package \cite{ARPREC}).
For more details on
the method and on the various techniques introduced in the
literature to fit the spectrum, see
\cite{Caf93,FMB03,GPS98,MBF05,PF07,PS98,Sh92,SSF83}.

\section{Van Dommelen and Shen's singularity}\label{VDSsing}

In this section we shall report the results of the singularity
tracking method in the case of the impulsively started disk. As we
have explained in the previous section, the method can be applied to
functions which are analytic. Strictly speaking there is no result
in the literature that ensures that Van Dommelen and Shen's datum is
analytic both in $x$ and in $Y$. In fact the result in \cite{LCS03}
ensures that, if the initial datum is analytic in $x$ and $C^2$ in
$Y$, then the solution will remain regular for a short time. However
we conjecture that, under the above hypotheses, the viscosity in $Y$
makes the solution, after a short time, analytic in $Y$. In fact the
solution of Prandtl's equations  can be put in the form \be
u(x,Y,t)=H(Y,t)*_{Y,t}\left[ -u\pardx u -v\pardY u +U\pardx U\right]
+IC+BC \; , \ee where $H$ is the heat kernel in $\mathbb{R}^+$,
where with $*$ we have denoted the convolution and where $IC$ and
$BC$ indicate terms which take into account  the initial and
boundary data (also these terms involve convolutions with the heat
kernel). Being the terms inside the square bracket $C^1$ in $Y$, it
should be a rather standard consequence of the properties of the
heat kernel the fact that the solution is analytic in $Y$ after a
short time. For the impulsively started disk a different problem is
represented by the discontinuity of the initial datum at the
boundary. However we believe that the regularizing properties of the
heat kernel make the solution $C^2$ (and therefore analytic). To
prove rigorously this statement is probably quite  involved,
as this would probably require an initial layer analysis.
The ultimate justification in the application of the singularity
tracking resides in the results (the spectra are exponentially
decaying within very good accuracy) and in their consistency with
the behavior of the solution in the physical space.

\begin{figure}
\begin{center}
\includegraphics[height=8.cm,width=11.5cm]{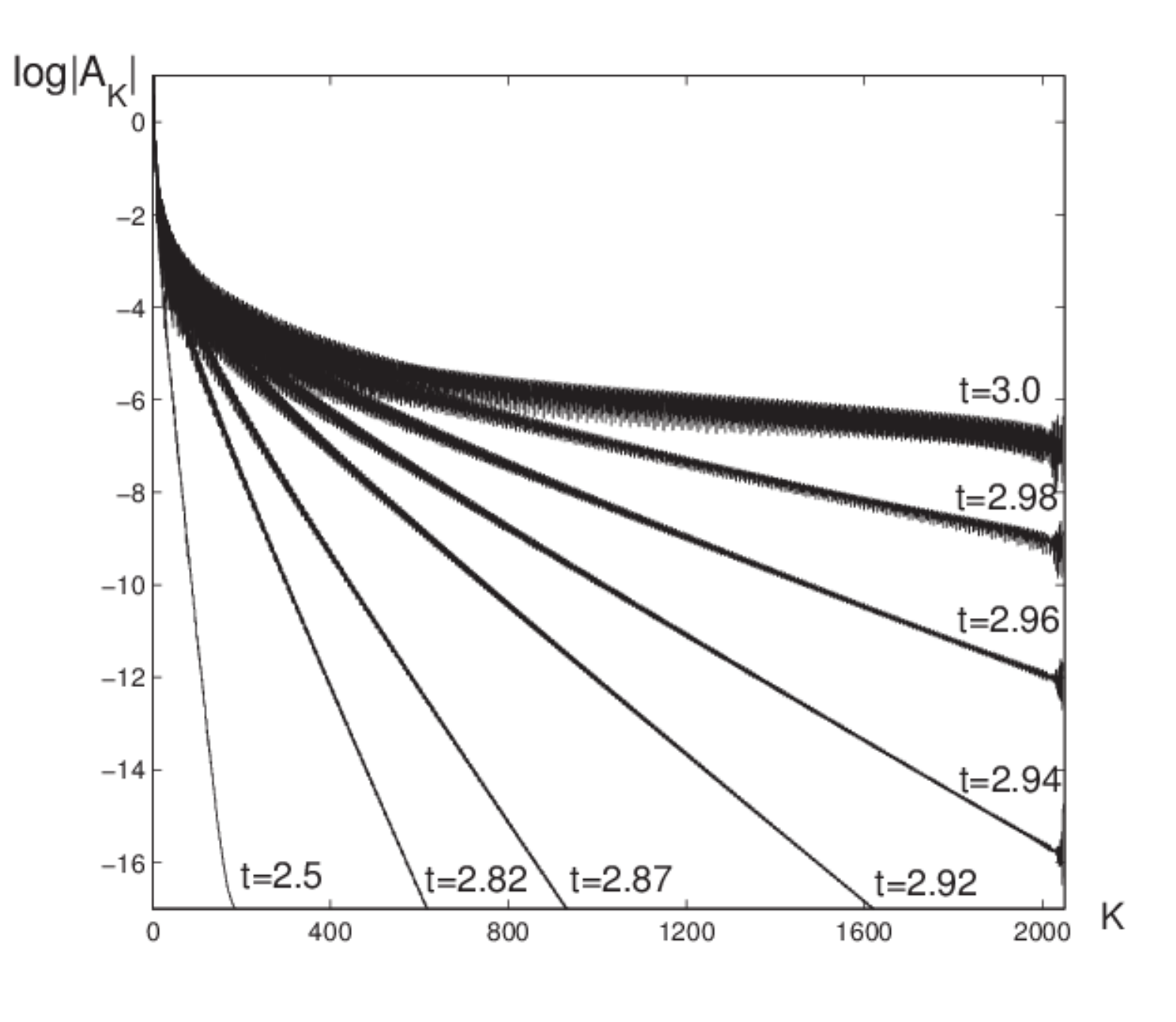}
\caption{{The behavior in time of the shell summed amplitude up
to the singularity time.}}\label{VDSshell}
\end{center}
\end{figure}

\begin{figure}
\begin{center}
\includegraphics[height=8.cm,width=11.5cm]{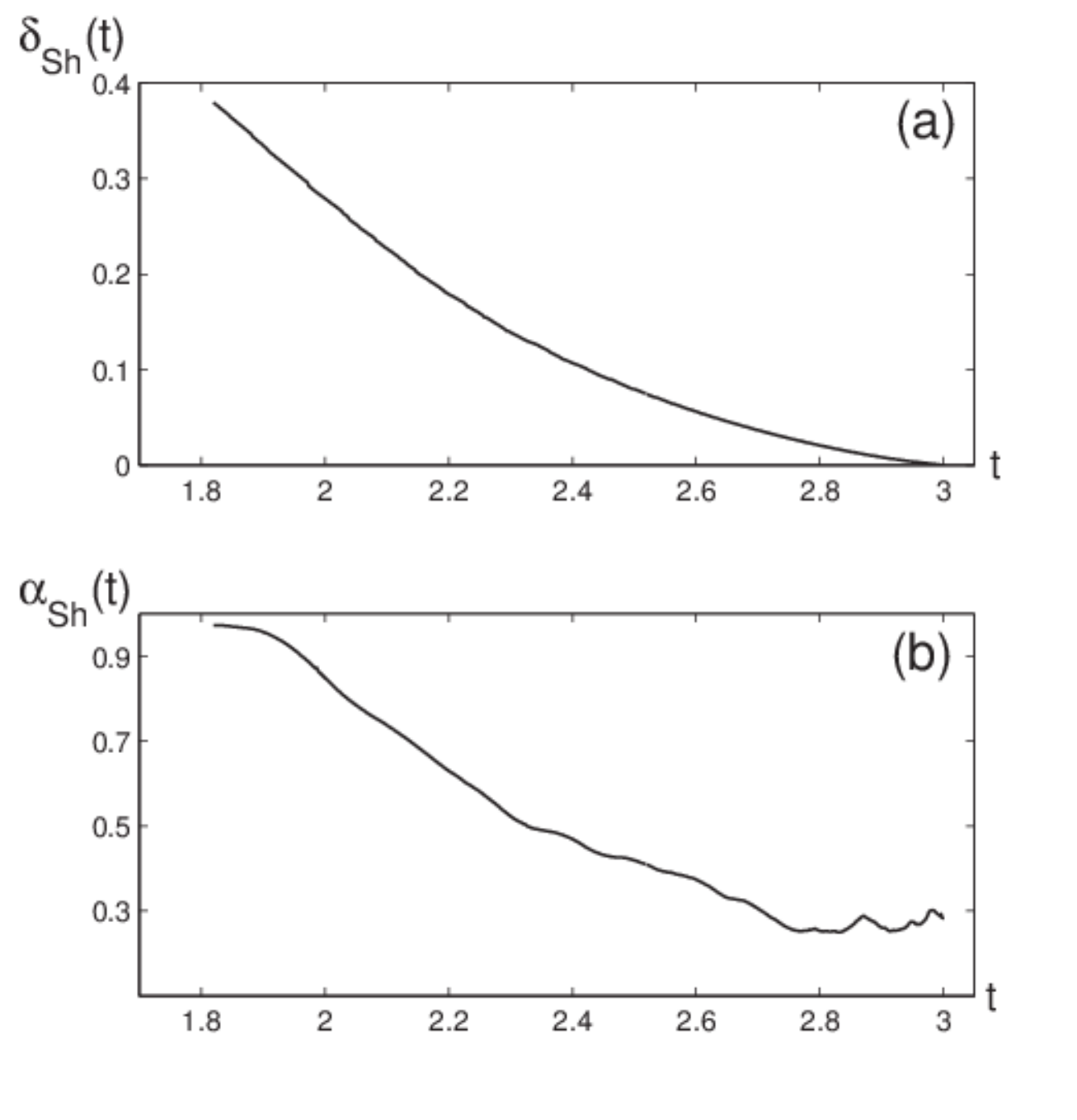}
\caption{\textit{(a)} The behavior in time of the width of the analyticity strip.
The singularity time is $t_c\approx 3.0$. \textit{(b)} The singularity is of cubic-root
type.}\label{VDSshellFITT}
\end{center}
\end{figure}

\begin{figure}
\begin{center}
\includegraphics[height=8.cm,width=11.5cm]{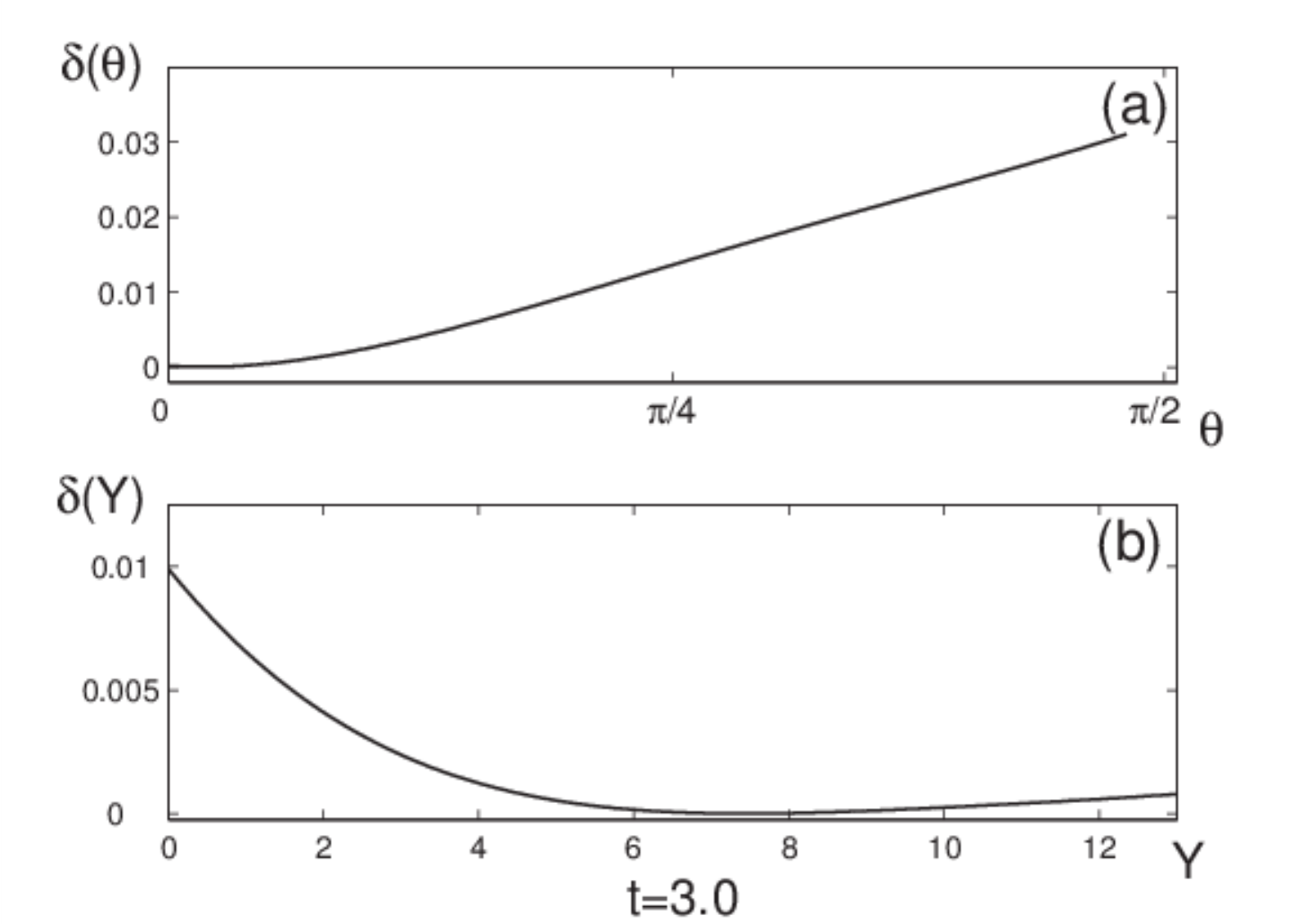}
\caption{\textit{(a)} The most singular direction is $\theta=0$.
\textit{(b)} If one estimates the $\delta$ in its dependence on the normal variable,
one finds that the location of the singularity is at $Y\approx 7$.}\label{VDSdeltaY}
\end{center}
\end{figure}

In Fig. \ref{VDSshell} we show the shell-summed amplitudes, where it is evident
the loss of exponential decay.
Fitting these amplitudes we get the results shown in Fig. \ref{VDSshellFITT},
where one can see that, at the critical time $t_c=3.0$, the solution loses analyticity
as a cubic-root singularity hits the real axis.
In Fig. \ref{VDSdeltaY}(a) we show the angular dependence of $\delta$ and
one can see that the most singular direction is $\theta=0$.
This result allows to treat the normal variable as a parameter.
At the bottom of the same figure we show the dependence of $\delta$ on  $Y$,
and one can see that  $\delta(Y)$ attains its  minimum at $Y\approx 7$.

\begin{figure}
\begin{center}
\includegraphics[height=8.cm,width=11.5cm]{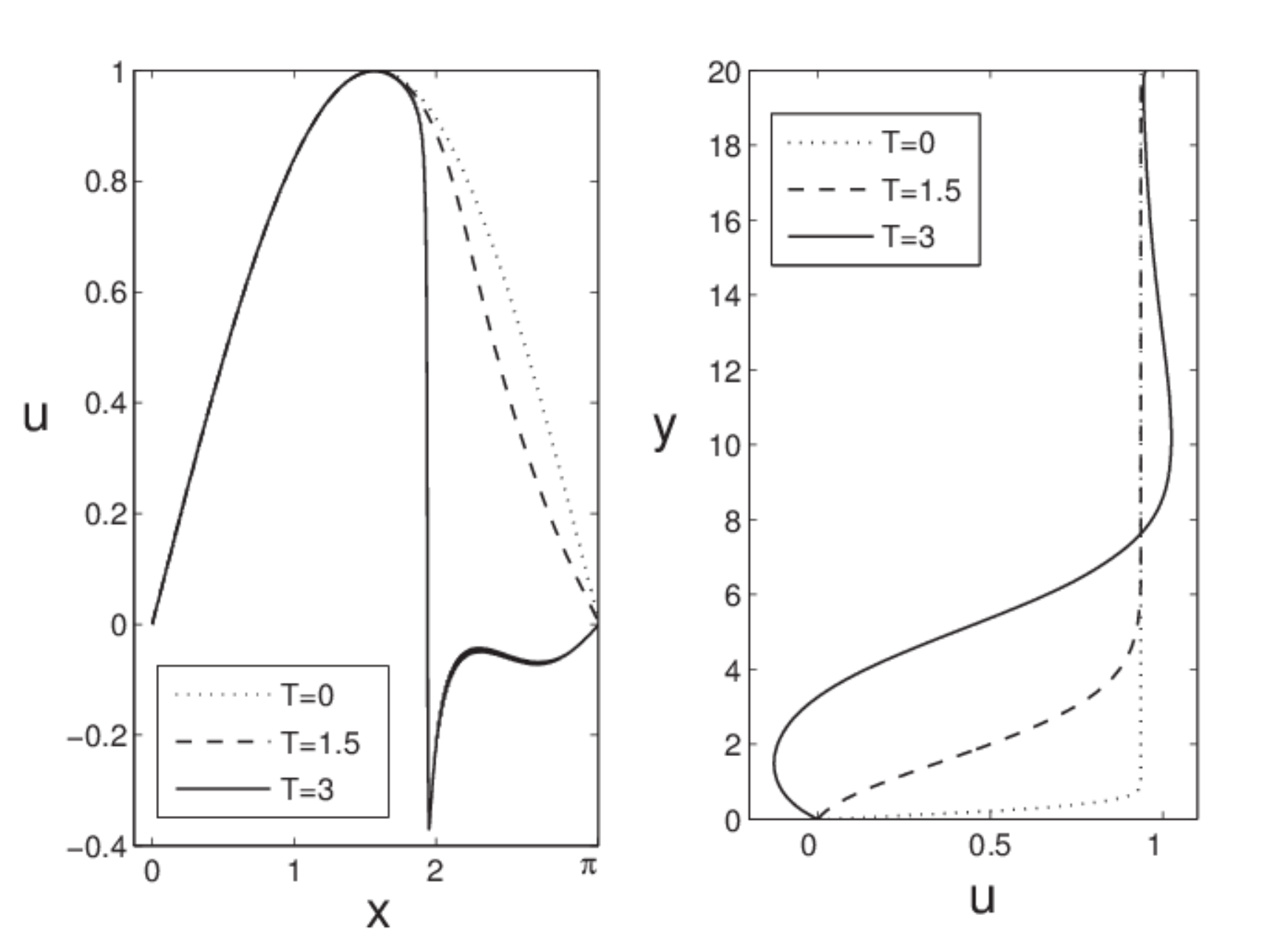}
\caption{The
solution of Prandtl's equation at the location $Y=7$ for VanDommelen and Shen
initial datum given by \eqref{P5}. To the
left one can follow the formation of a shock at $x\approx 1.94$ at the
location $Y=7$. The right figure   shows the tangential
velocity at the singularity location along the boundary
layer.}\label{VDSfig}
\end{center}
\end{figure}

\begin{figure}
\begin{center}
\includegraphics[height=8.cm,width=11.5cm]{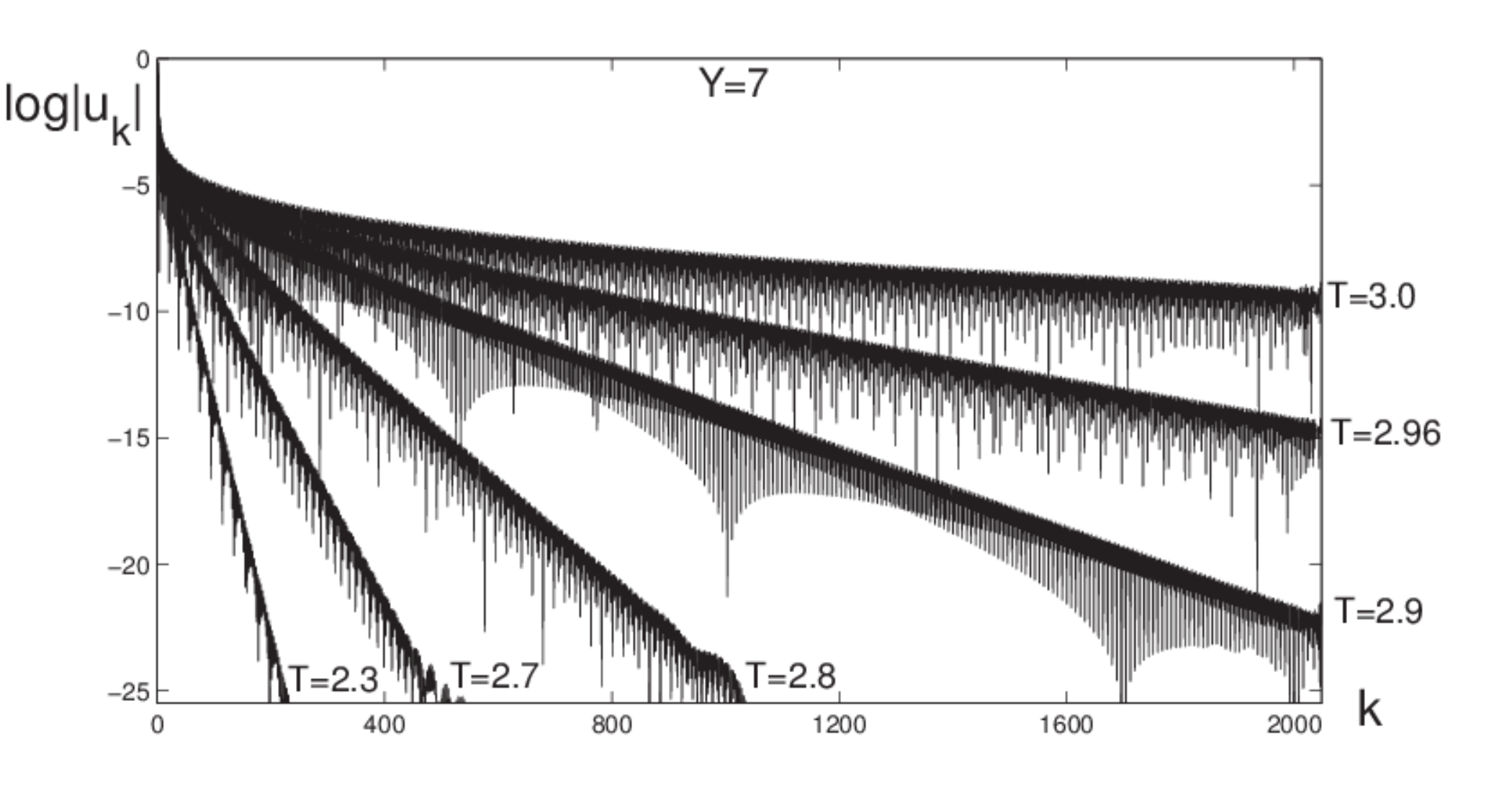}
\caption{The spectrum of the
solution of Prandtl's equation at the location $Y=7$ with initial
datum given by \eqref{P5}.}\label{VDSspec}
\end{center}
\end{figure}

\begin{figure}
\begin{center}
\includegraphics[height=8.cm,width=11.5cm]{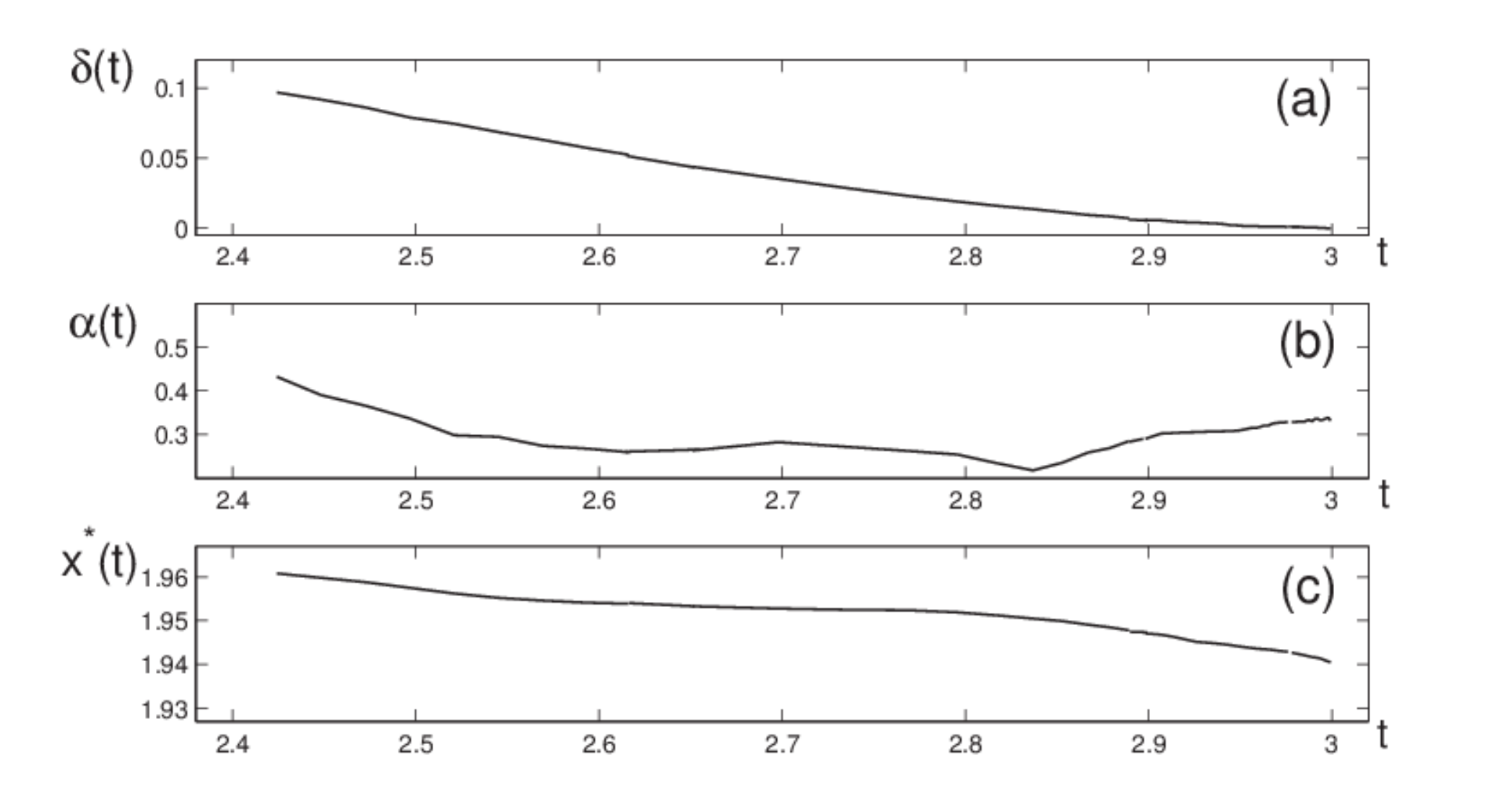}
\caption{The results of the
singularity tracking at $Y=7$. One can see that at $t\approx 3$ the strip
of analyticity shrinks to zero as the result of a cubic--root
singularity hitting the real axis. The location of the shock in
Fig. \ref{VDSfig} and the real coordinate of the singularity $x^*$
at the singularity time coincide.}\label{VDSfit}
\end{center}
\end{figure}

In Fig. \ref{VDSfig} we show the solution, at the singularity time,
at the location $Y=7$ and in Fig. \ref{VDSspec} it is show the behaviour in time of the spectrum at location $Y=7$.
In Fig. \ref{VDSfit} one can see the results of the singularity
tracking method for the initial datum \eqref{P5} at the location
$Y=7$ obtained using the mixed spectral-finite difference numerical scheme.

All these results are consistent with those obtained with the fully spectral
method.
In Fig.\ref{VDSfit}(c) one can see that the real tangential location
of the singularity $x^*$, found with a study of the oscillatory behavior of the spectrum depurated by the exponential and algebraic decay, is consistent with the location of the shock shown in Fig. \ref{VDSspec}.

\section{Early singularity formation}\label{ESF}
\setcounter{equation}{0}

In the rest os this paper we shall denote
by $\mathcal{U}\equiv u^{VDS}(x,Y,t=1.5)$
the Van Dommelen and Shen solution  (i.e. the solution of \eqref{P1}--\eqref{P4}
with initial datum \eqref{P5}) at the time $t=1.5$, well before the critical time.
In this section we shall \textit{perturb} this datum adding an analytic (in $x$) function
that has a singularity at distance $\delta_0$ from the real axis.
We shall show that the addition of this perturbation accelerates
the formation of the singularity.

Let us introduce the family of real functions:
\begin{equation}\label{perturb}
U_{\delta_0}(x)=\sum_{2\leq|k|\leq K/2}
-i\frac{k}{|k|}\frac{e^{-\delta_0 |k|}}{|k|^{5/3}}  \cos{(x^*
k)}e^{ikx},
\end{equation}
The above functions are analytic in a strip of the complex plane of width $\delta_0$ and
have a $2/3$--singularity at distance $\delta_0$ from the real axis.
Moreover the fact that the spectrum oscillates as $ \cos{(x^* k)}$ means that
we are considering a dipole singularity, i.e. two $5/3$--singularities of opposite
sign located at $x^*\pm\varepsilon+i\delta$ of strength $1/\varepsilon$ in the limit
$\varepsilon\rightarrow 0$.

We now consider the following family of initial data for Prandtl's
equations \eqref{P1}--\eqref{P4}. \be u_0(x,Y)=\mathcal{U}+\sigma
U_{\delta_0} f(Y)\; , \label{iniperturb} \ee where $f(y)$ is a
$C^2$ function of the form: \be f(y)=\left\{
\begin{array}{ll}
y^2\exp{(-y)} & \mbox{for}  \; 0\leq y\leq 2\\
4e^{-2} \quad &\mbox{for}  \;2\leq y\leq 3 \\
4e^{-2}\exp{\left[(-(y-3)/1.5)^3\right]} &\mbox{for}  \;  3\leq
y\leq 20
\end{array}\right.
\ee
This means that we are perturbing the VDS datum at $t=1.5$
placing a singularity at distance $\delta_0$ from the real axis.
Moreover the role of the function $f(Y)$ is to make the strength
of the singularity zero at the boundary and exponentially decaying
far away from the the boundary. The maximum of the perturbation
when $2\leq Y\leq 3$.
The role of $\sigma$ is to tune the strength of the perturbation
of the VDS datum. In all the calculations we shall present we have
chosen $\sigma=6.6$. Moreover we have taken $x^*=2.17$ which is
the real coordinate  of the VDS singularity at $t=1.5$.

\begin{figure}[]
\begin{center}
\includegraphics[height=7.5cm,width=9.5cm]{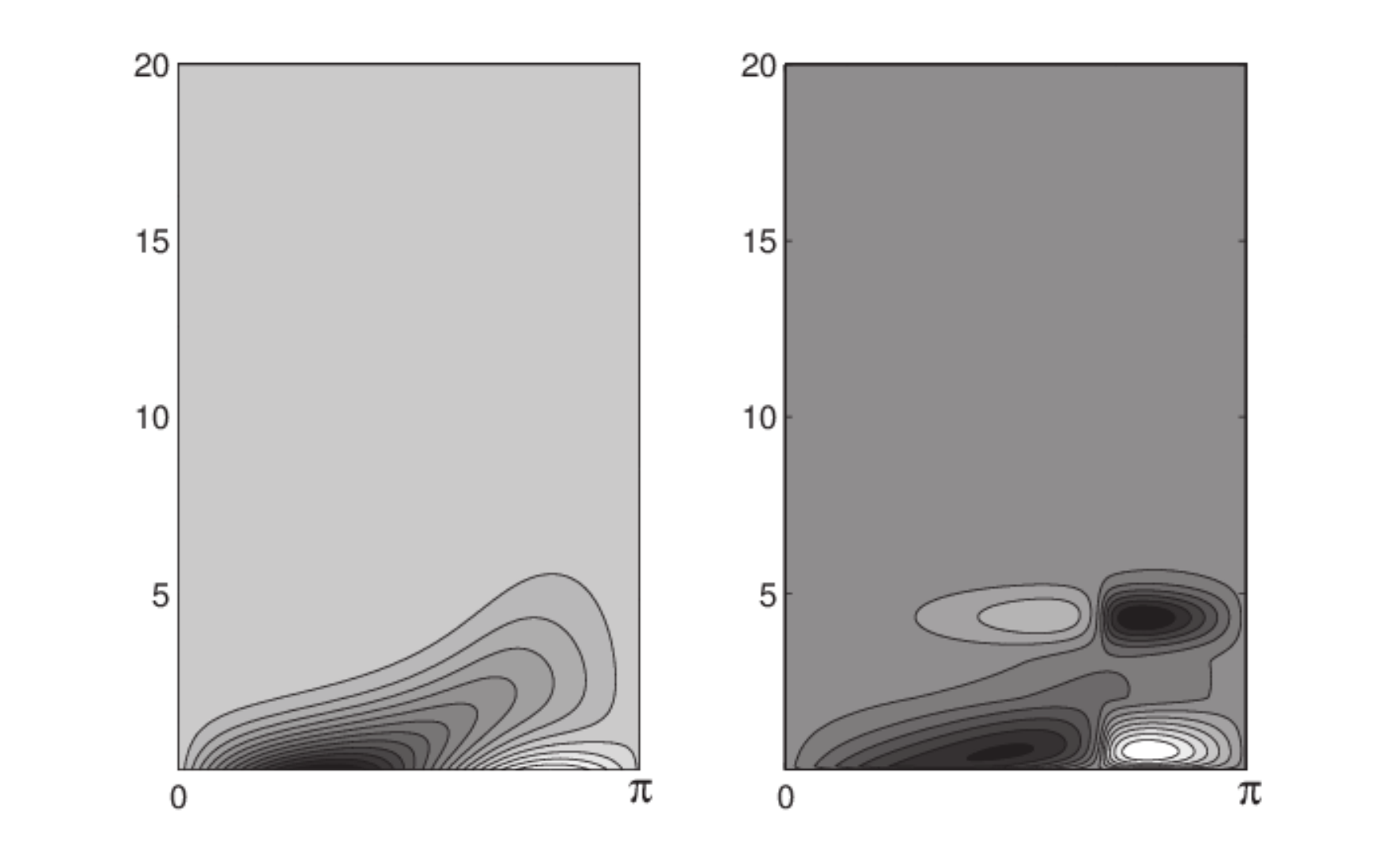}
\caption{  {To the left the
vorticity   of the VDS solution at $t=1.5$. To the right the
vorticity of the datum \eqref{iniperturb} with $\delta_0=0.025$.
}}\label{UCH1ps_2}
\end{center}
\end{figure}

\begin{figure}[]
\begin{center}
\includegraphics[height=7.5cm,width=13.5cm]{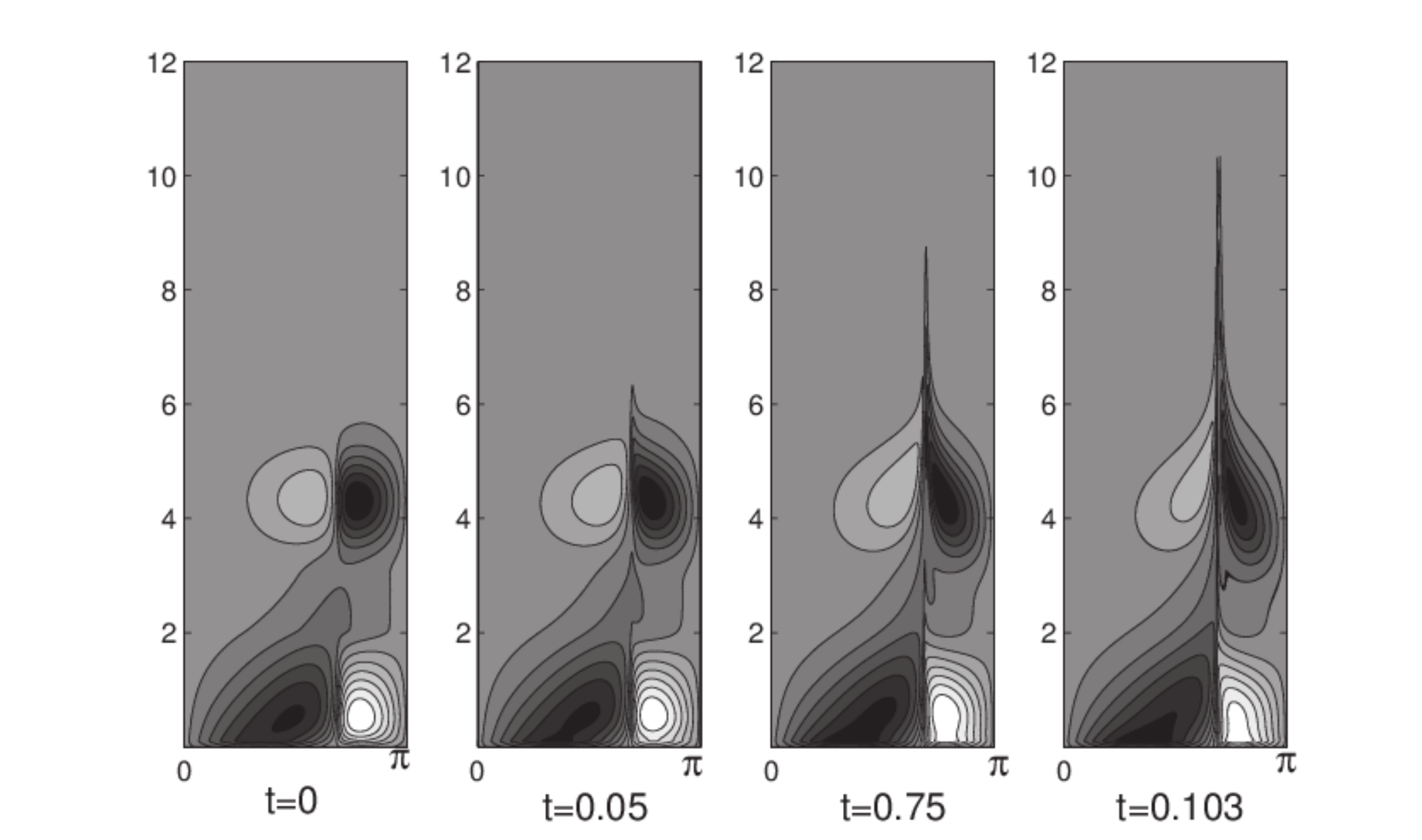}
\caption{  {The vorticity
$-\partial_Y u$ for Prandtl's equations with initial datum
\eqref{iniperturb} and $\delta_0=0.025$, at different times.
}}\label{UCH1ps_3}
\end{center}
\end{figure}

In Fig. \ref{UCH1ps_2} we compare the vorticity (defined as
$-\partial_Y u$) of $\mathcal{U}$ (to the left) with the vorticity
of the initial datum \eqref{iniperturb} with $\delta_0=0.025$ (to
the right). From Fig. \ref{UCH1ps_2} it is evident that the effect
of the perturbation $\sigma U_{\delta_0}f(Y)$ is to place a
couple of counter rotating vortices in the boundary layer. One can
see in Fig. \ref{UCH1ps_3} how these two vortices speed up the
process of the vorticity stretching
and the process of the
singularity formation for Prandtl's equations.

It is important to notice that the family of functions
\eqref{perturb} when $\delta_0\geq 0$ is uniformly bounded in the
norm  $H^1$ (actually in $H^s$ when $s<7/6$).

We now solve Prandtl's equations \eqref{P1}--\eqref{P4} with initial datum given by
\eqref{iniperturb} with $\delta_0=0.025$.
The calculations we shall present here have been performed using a fully
spectral method with resolution $2048^2$.

\begin{figure}[]
\begin{center}
\includegraphics[height=7.5cm,width=9.5cm]{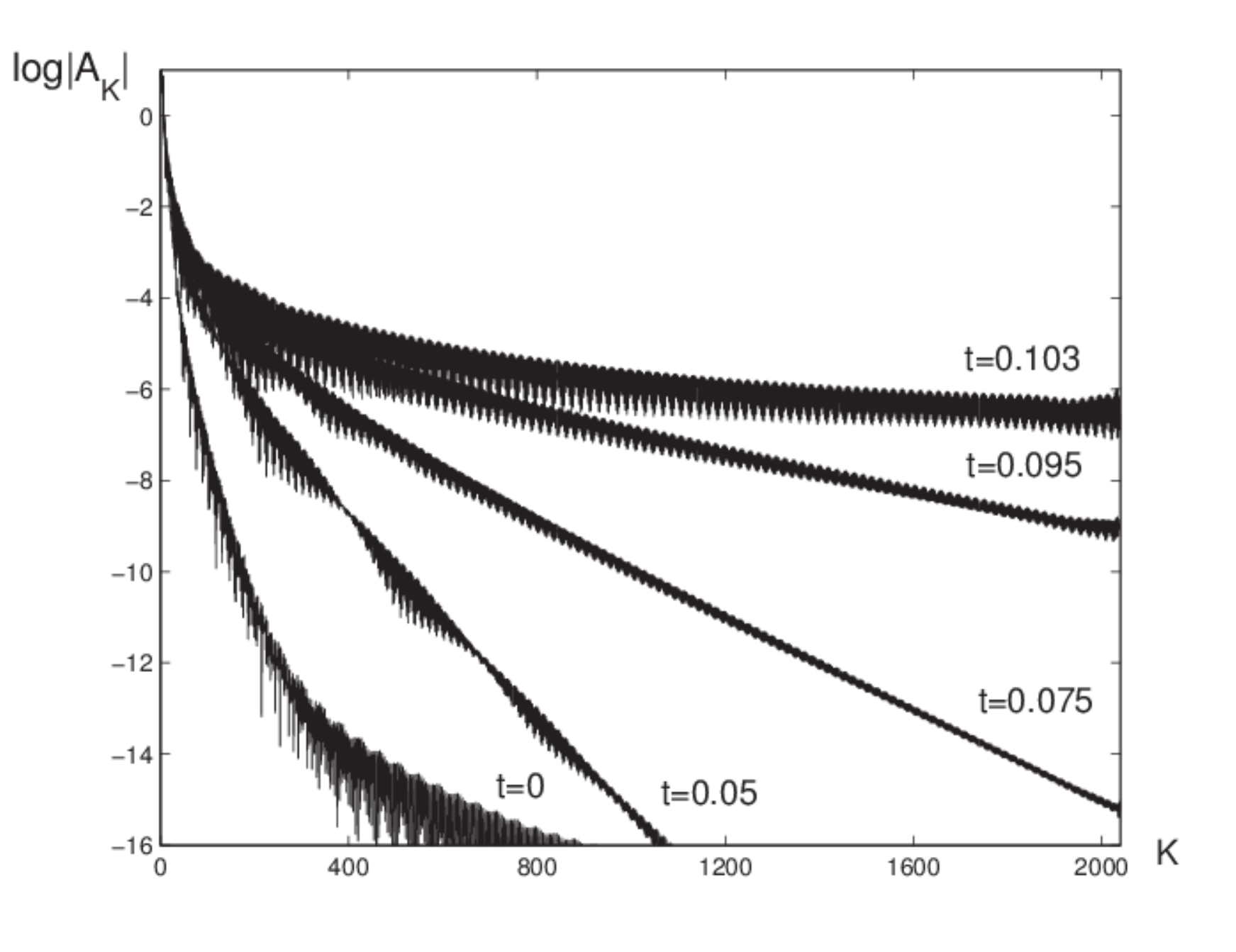}
\caption{  {The behavior in time of the shell summed amplitude up
to the singularity time.
}}\label{shell_VDSpert025}
\end{center}
\end{figure}

\begin{figure}[]
\begin{center}
\includegraphics[height=8.5cm,width=9.cm]{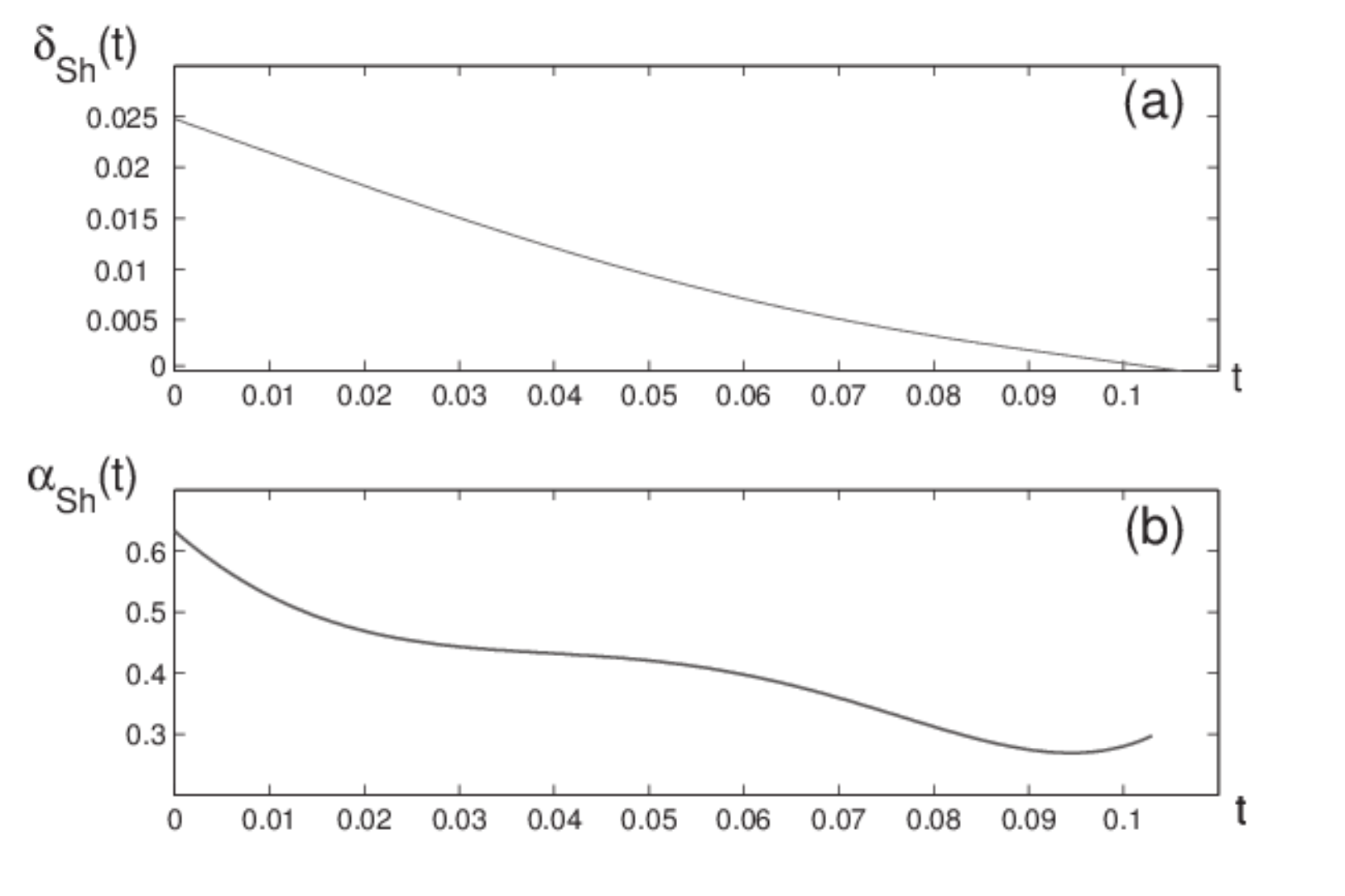}
\caption{ \textit{(a)} The behavior in time of the width of the analyticity strip.
The singularity time is $t_c\approx 0.103$. \textit{(b)} The singularity is of cubic-root
type.
}\label{shell_fitting_delta_alpha_VDSpert025}
\end{center}
\end{figure}

\begin{figure}[]
\begin{center}
\includegraphics[height=7.5cm,width=8.5cm]{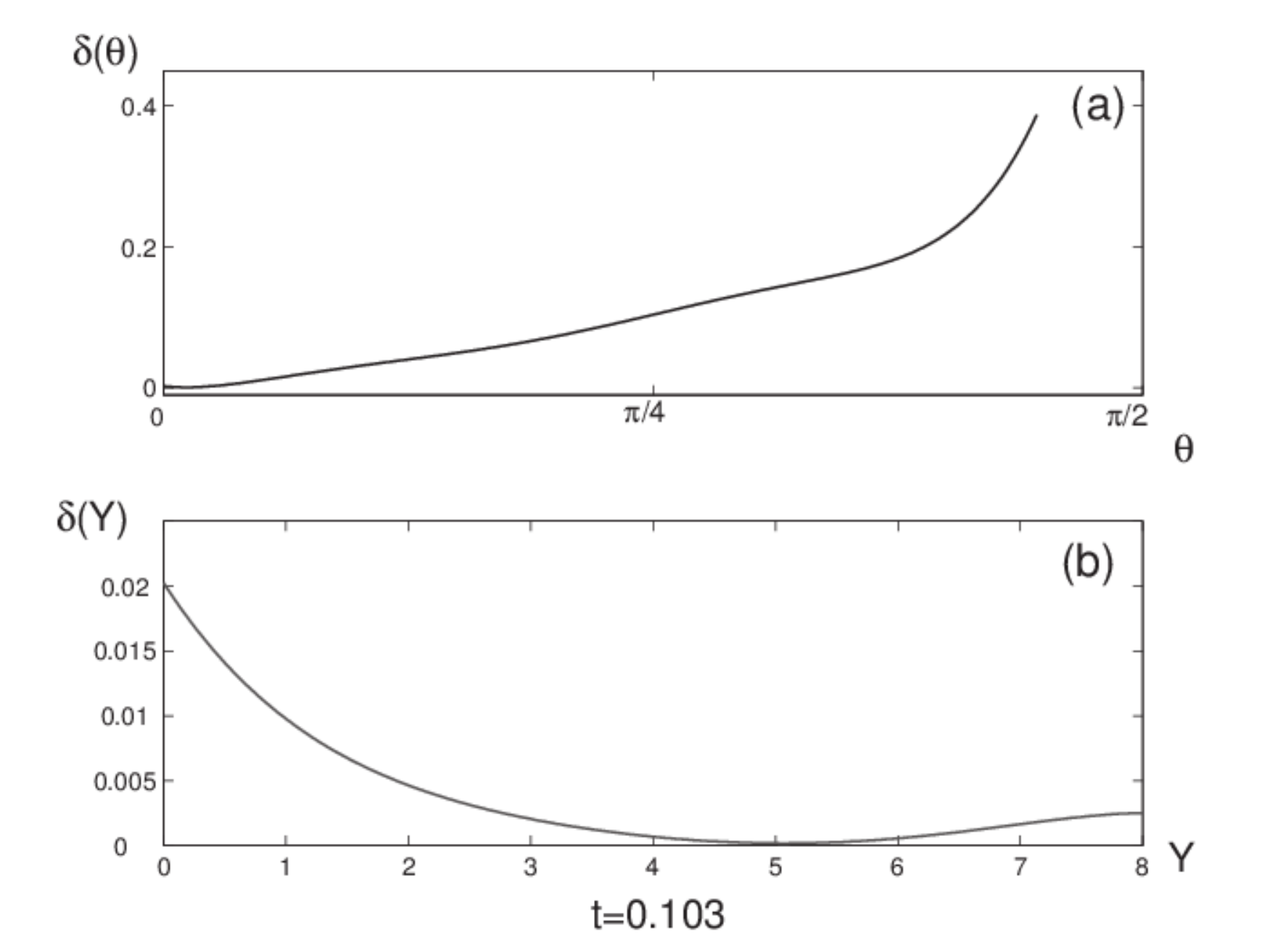}
\caption{ \textit{(a)} As in the case of the Van Dommelen and Shen singularity,
the most singular direction is $\theta=0$.
\textit{(b)} If one estimates the $\delta$ in its dependence on the normal variable,
one finds that the location of the singularity is at $Y\approx 5$.
}\label{fitt_VDSpert_025_delta_Y_theta}
\end{center}
\end{figure}

In Fig. \ref{shell_VDSpert025} we show the shell-summed amplitudes, where it is evident
the loss of exponential decay.
Fitting these amplitudes we get the results shown in Fig. \ref{shell_fitting_delta_alpha_VDSpert025},
where one can see that, at time $t\approx 0.103$, the solution loses analyticity
as a cubic-root singularity hits the real axis.
In Fig. \ref{fitt_VDSpert_025_delta_Y_theta}(a) we show the angular dependence of $\delta$ and
one can see that the most singular direction is $\theta=0$.
This result allows to treat the normal variable as a parameter.
At the bottom of the same figure we show the dependence of $\delta$ on  $Y$,
and one can see that  $\delta(Y)$ attains its  minimum at $Y\approx 5$.

\begin{figure}[ht]
\begin{center}
\includegraphics[height=7.5cm,width=10.5cm]{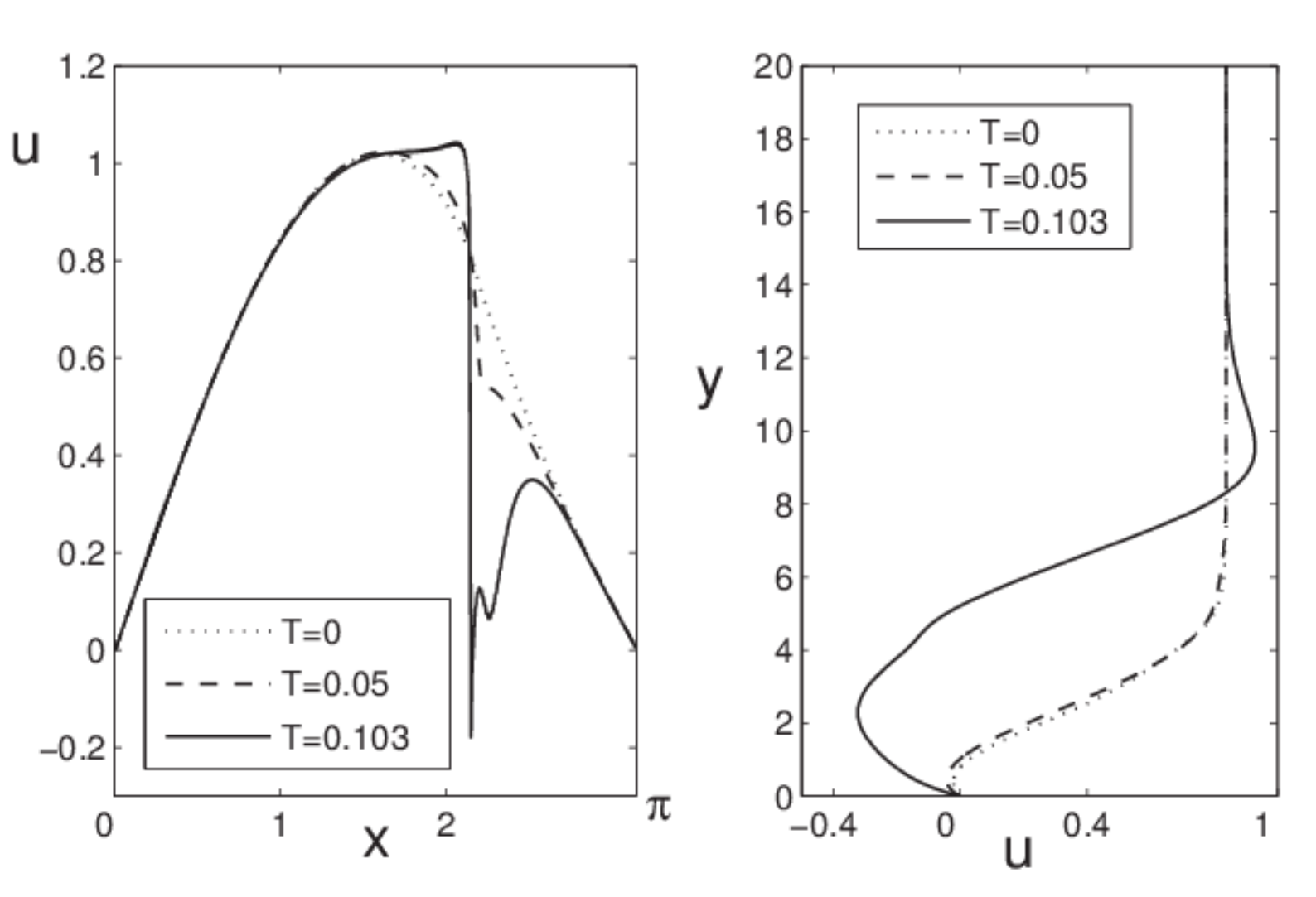}
\caption{ The
solution of Prandtl's equation at the location $Y=5$ with initial
datum given by \eqref{iniperturb} with $\delta_0=0.025$. To the
left one can follow the formation of a shock at $x\approx 2.14$ at the
location $Y=5$. The right figure   shows the tangential
velocity at the singularity location along the boundary
layer.}\label{deltap025}
\end{center}
\end{figure}

\begin{figure}[ht]
\begin{center}
\includegraphics[height=6.5cm,width=13.5cm]{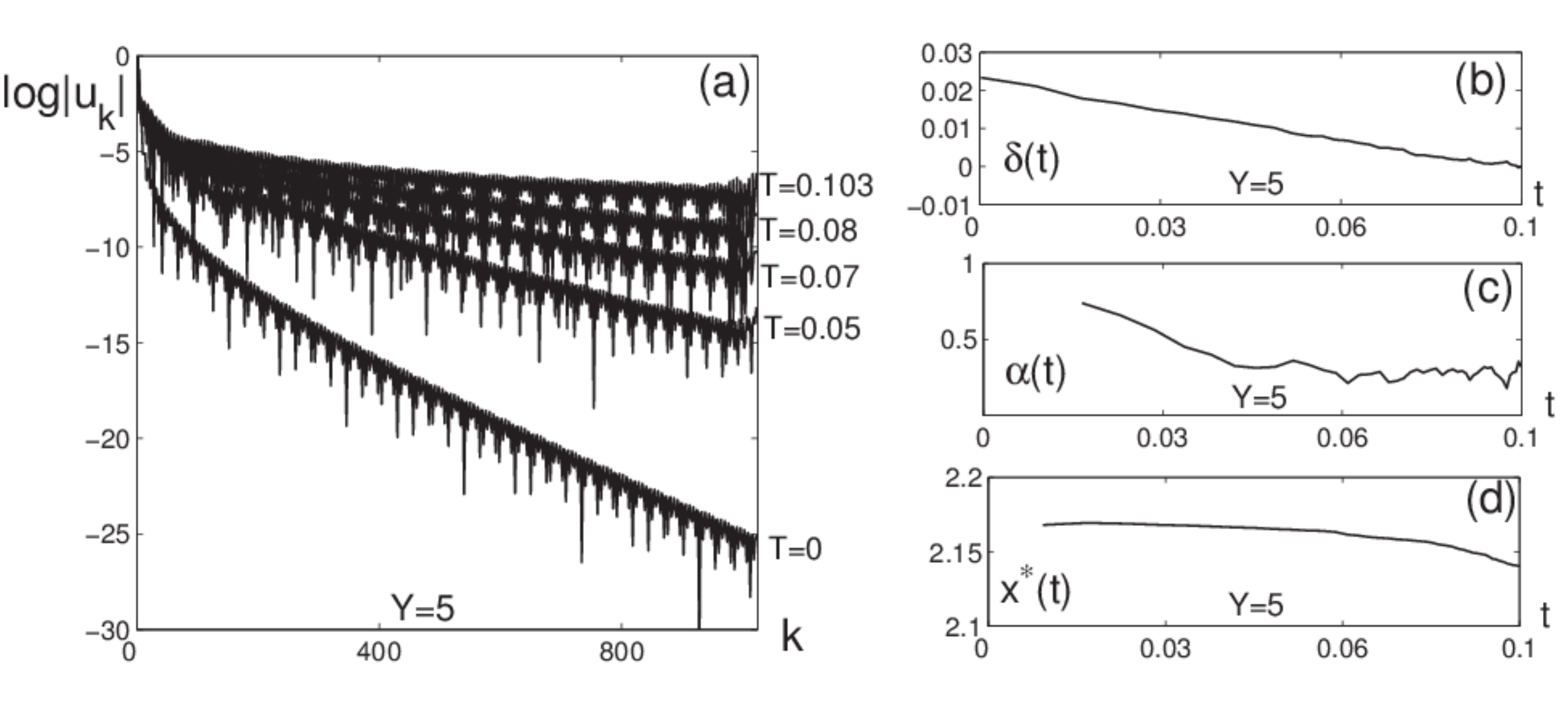}
\caption{
\textit{(a)} The spectrum of the solution of Prandtl's equation at the location
$Y=5$ with initial datum given by \eqref{iniperturb} with
$\delta_0=0.025$.
\textit{(b)} and \textit{(c)} The results on the singularity time and on the
type of the singularity, obtained fitting the spectrum at $Y=5$,
are consistent with the results obtained using the shell-summed amplitudes.
\textit{(d)}
The location of the shock in
Fig.\ref{deltap025} and the real coordinate $x^*$ of the
singularity, at the blow-up time, coincide.}\label{deltap025bis}
\end{center}
\end{figure}

In Fig. \ref{deltap025} we show the solution, at the singularity time,
at the location $Y=5$.
In Fig.\ref{deltap025bis} we show the results obtained using the mixed
spectral-finite difference numerical scheme at the location $Y=5$.
All these results are consistent with those obtained with the fully spectral
method.
In Fig.\ref{deltap025bis}(d) one can see that the real tangential location
of the singularity $x^*$, found with a study of the oscillatory behavior of the spectrum
depurated by the exponential and algebraic decay, is consistent with the location of the shock shown in
Fig. \ref{deltap025}(a).


\section{A case study: Burger's equation}\label{casestudy}
\setcounter{equation}{0}

\begin{figure}[ht]
\begin{center}
\includegraphics[height=7.5cm,width=10.5cm]{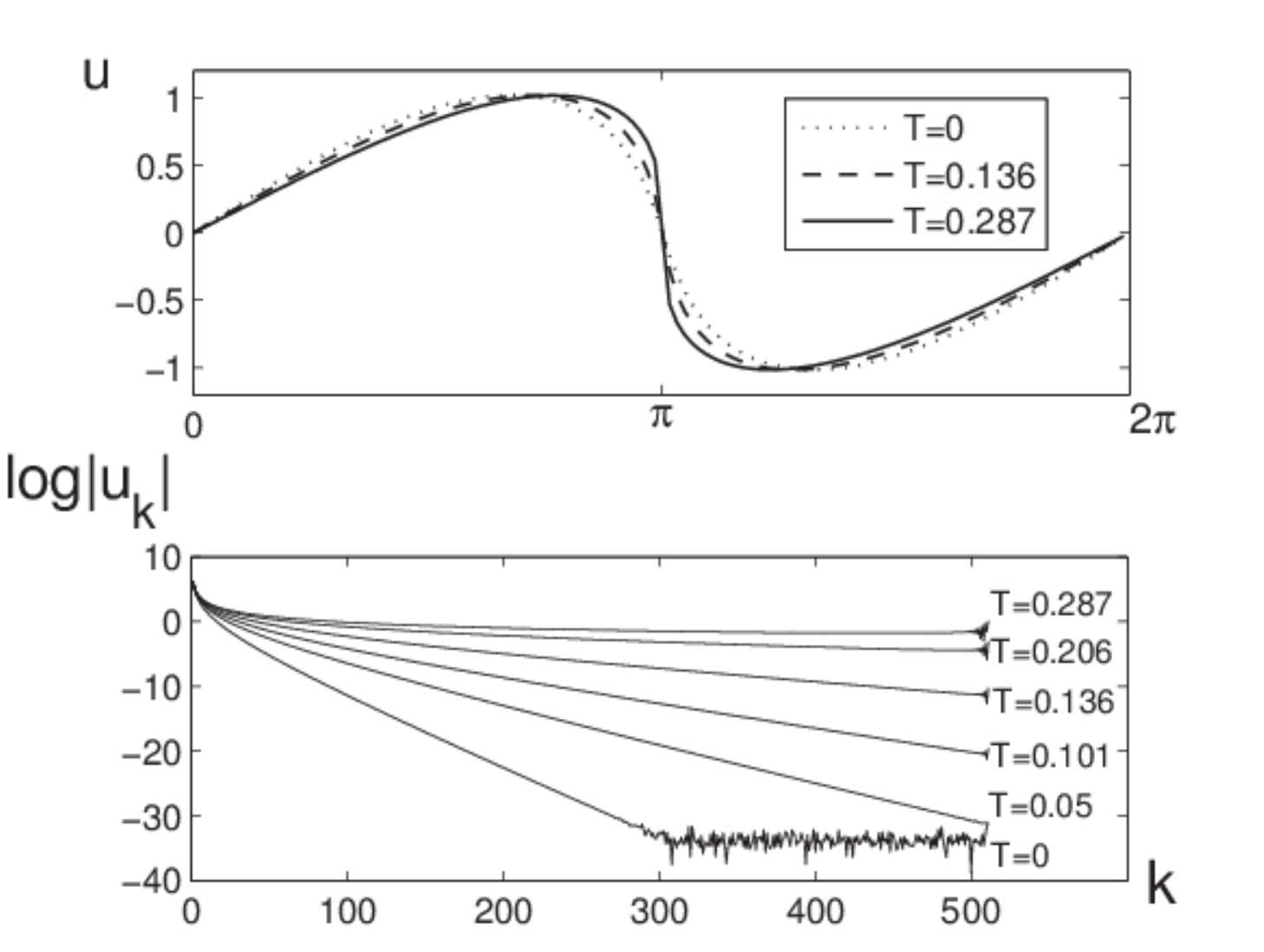}
\caption{The behavior of the
solution and the spectrum of Burger's equation with initial
datum \eqref{Burginidata} with $\delta_0=0.1$. A singularity in
the form of a weak (see also Fig. \ref{burg53alphadeltatimes})
shock forms at $t\approx 0.287$. }\label{burger53solspec}
\end{center}
\end{figure}

We now consider Burger's equation:
$$
\pardt u+u\pardx u=0 \; ,
$$
with initial datum: \be u(x,t=0)=\sin{(x)}+U_{\delta_0} \; ,
\label{Burginidata} \ee with $U_{\delta_0}$ given by
\eqref{perturb} where we have chosen $x^*=\pi$. At the top of Fig.
\ref{burger53solspec} we show the behavior of the solution at
various time for $\delta_0=0.1$. At the bottom one can see  the
behavior of the Fourier spectrum at various times. The loss of
exponential decay of the spectrum is apparent. Moreover, the fact
that the spectrum does not oscillate, clearly indicates that the
real coordinate of the singularity remains $x^*(t)=\pi$.

\begin{figure}[ht]
\begin{center}
\includegraphics[height=6.5cm,width=10.5cm]{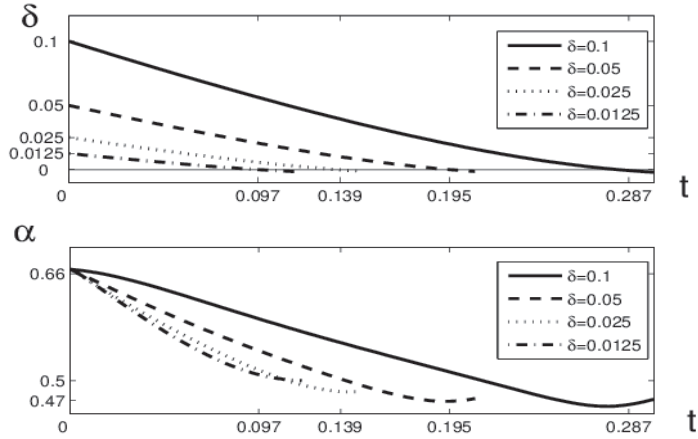}
\caption{At
the top the  distances of the singularity from the real axis
for different initial $\delta_0$s. At the bottom one can see
that, at the singularity times, all the singularities have evolved
in square--root singularities.}\label{burg53alphadeltatimes_1}
\end{center}
\end{figure}

At the top of Fig. \ref{burg53alphadeltatimes_1}  we are representing the shrinking in time
of the analyticity strip for various $\delta_0$.
At the bottom the evolution in time of the singularity type for different $\delta_0$.
One can see that at the respective singularity times, the solutions form a square root
singularity.

\begin{figure}[ht]
\begin{center}
\includegraphics[height=4.5cm,width=10.5cm]{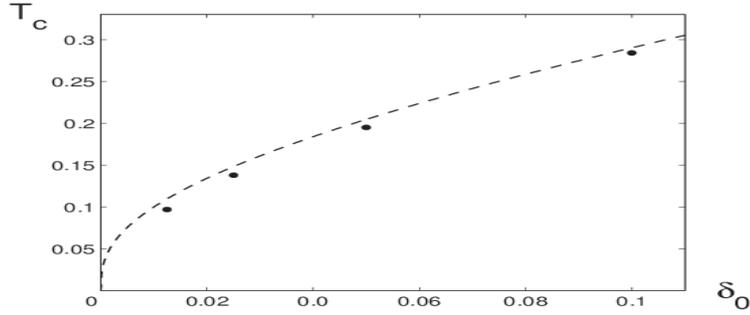}
\caption{The singularity times
for different $\delta_0$. The singularity time goes to zero when
the $\delta_0$ goes to zero. }\label{burg53alphadeltatimes_2}
\end{center}
\end{figure}

In Fig. \ref{burg53alphadeltatimes_2} we compare the computed
singularity times for different initial $\delta_0$
with the theoretical values of the critical times (the dashed line)
In fact, for Burger's equation, the singularity time is related to
the inverse of $\sup_x|\pardx u_0|$.
For the initial data $u_0$ given by \eqref{Burginidata}, one has:
\begin{equation}
\sup_x|\pardx u_0|=1+\sum_{k>1}\frac{e^{-\delta_0 k}}{k^{2/3}}.
\end{equation}
Introducing the polylogarithm   $Li_{s}(z)=\sum_{k\geq
1}\frac{z^{ k}}{k^{s}}$ (see \cite{AS72}), one can write the above
formula as
\begin{equation}
\sup_x|\pardx
u_0|=1-e^{-\delta_0}+Li_{\frac{2}{3}}(e^{-\delta_0}).
\end{equation}
The polylogarithm  $Li_{s}(z)$ with $s=2/3$ is analytic for $|z|<1$
and divergent at $z=1$ This  means that in the limit $\delta_0
\rightarrow 0$, the derivative of the initial data $u_0$ becomes
arbitrarily large and therefore the singularity time arbitrarily
short.
Therefore in the class of initial data
\eqref{Burginidata} (which, as we previously noticed is uniformly
bounded in $H^1$ when $\delta_0\geq 0$) the singularity time can
be made arbitrarily short. The fact that the singularity evolves
in a square-root singularity means that the solution, at the
singularity time, it is not in $H^1$.

\begin{figure}[ht]
\begin{center}
\includegraphics[height=5.5cm,width=13.5cm]{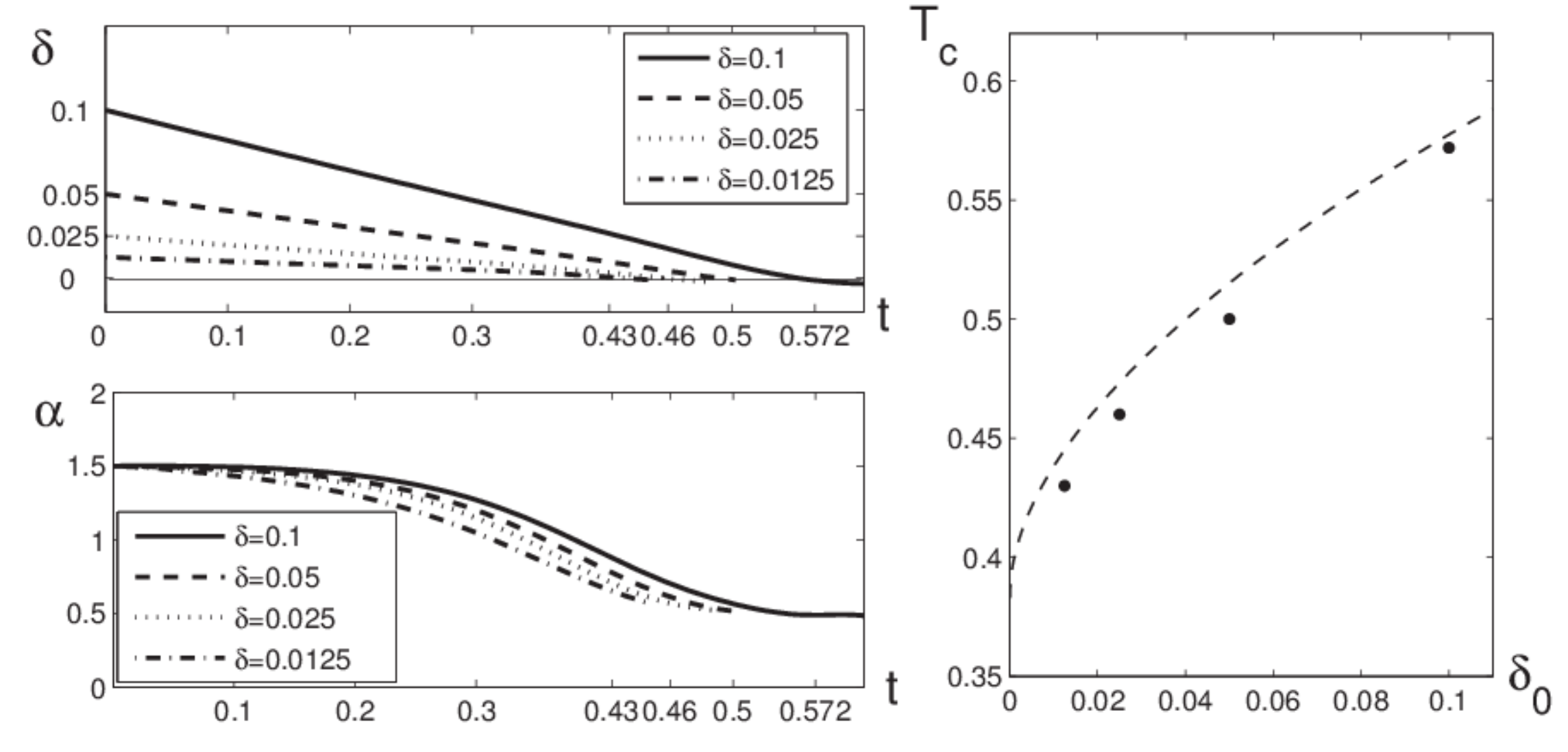}
\caption{If one
considers the class of initial data \eqref{pertur52} one can see
in the right figure that, however small the $\delta_0$ is taken,
the time for which the solution becomes singular does not converge
to zero. At the singularity time the solutions develop a
square--root singularity (bottom left  figure).
}\label{burg52alphadeltatimes}
\end{center}
\end{figure}

The situation is in fact different for the class of initial data:
\be u_0(x,y)=\sin{(x)}-\sum_{2 \leq|k|\leq K/2}
2i\frac{k}{|k|}\frac{e^{-\delta_0 |k|}}{|k|^{5/2}}  \cos{(x^*
k)}e^{ikx}, \label{pertur52} \ee
These data, due to the different
algebraic behavior of the spectrum (the complex singularities are
in fact weaker), they are uniformly bounded in $H^{3/2}$ where the
problem is well posed. This can be seen in Fig.
\ref{burg52alphadeltatimes} where one can see that these data
still develop a square--root singularity, but the singularity time
does not converge to zero when $\delta_0 $ goes to zero.

\section{The singularity times for Prandtl's equation}\label{Discussion}
\setcounter{equation}{0}

\begin{figure}[ht]
\begin{center}
\includegraphics[height=8.5cm,width=12.5cm]{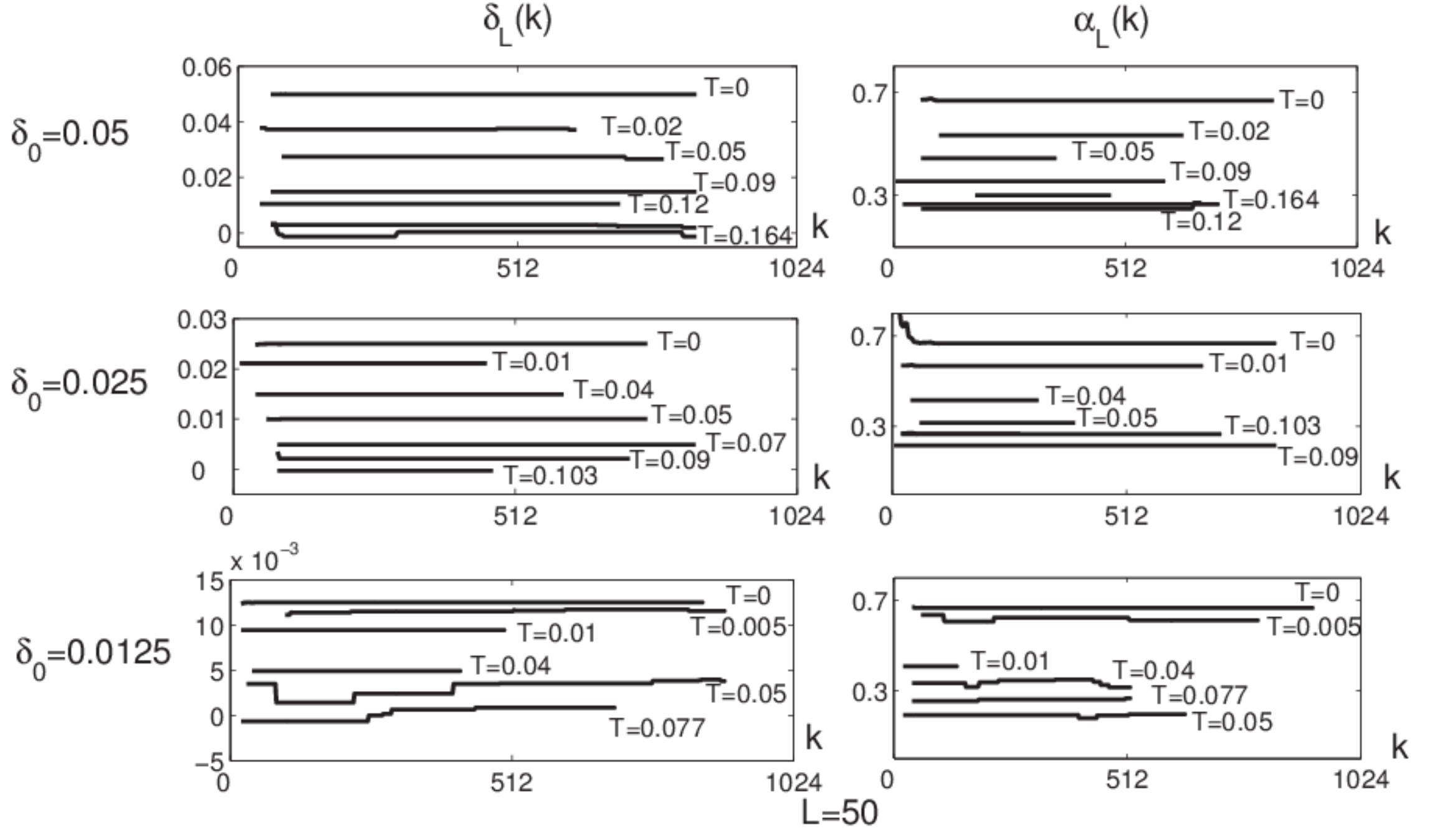}
\caption{{The results of the
sliding fitting of the spectrum of the solution of Prandtl's equation, at
the location Y=5, with initial datum \eqref{iniperturb} and with different
$\delta_0$. The left figures  show the width of the
analyticity strip as it goes to zero at the singularity times.
To the right the algebraic classification of the singularities.
At the singularity time they  are all cubic--root singularities.}
}\label{prandtlSliding53}
\end{center}
\end{figure}

We now solve the Prandtl equations with the initial data
\eqref{perturb} for different $\delta_0$.
To give more reliability to our results we have tracked the singularities
using  also the sliding fitting technique.
Instead of fitting the whole spectrum, one fits
(using the asymptotic formula \eqref{asyspectrum})
the spectrum  starting from the mode $u_k$ up to the mode $u_{k+L}$.
If the results (i.e. the $\delta$s and the $\alpha$s) are relatively
insensitive to the $k$, this can be considered as an evidence
of the genuineness of the results.
In the calculations we show in Fig.
\ref{prandtlSliding53} we have chosen $L=50$.
Similar results have been obtained with different $L$.
For other instances where the sliding fitting has been used,
we refer to \cite{Caf93,CBT99,DLSS06,GPS98,PS98} and to reference therein.

\begin{figure}[ht]
\begin{center}
\includegraphics[height=6.5cm,width=9.5cm]{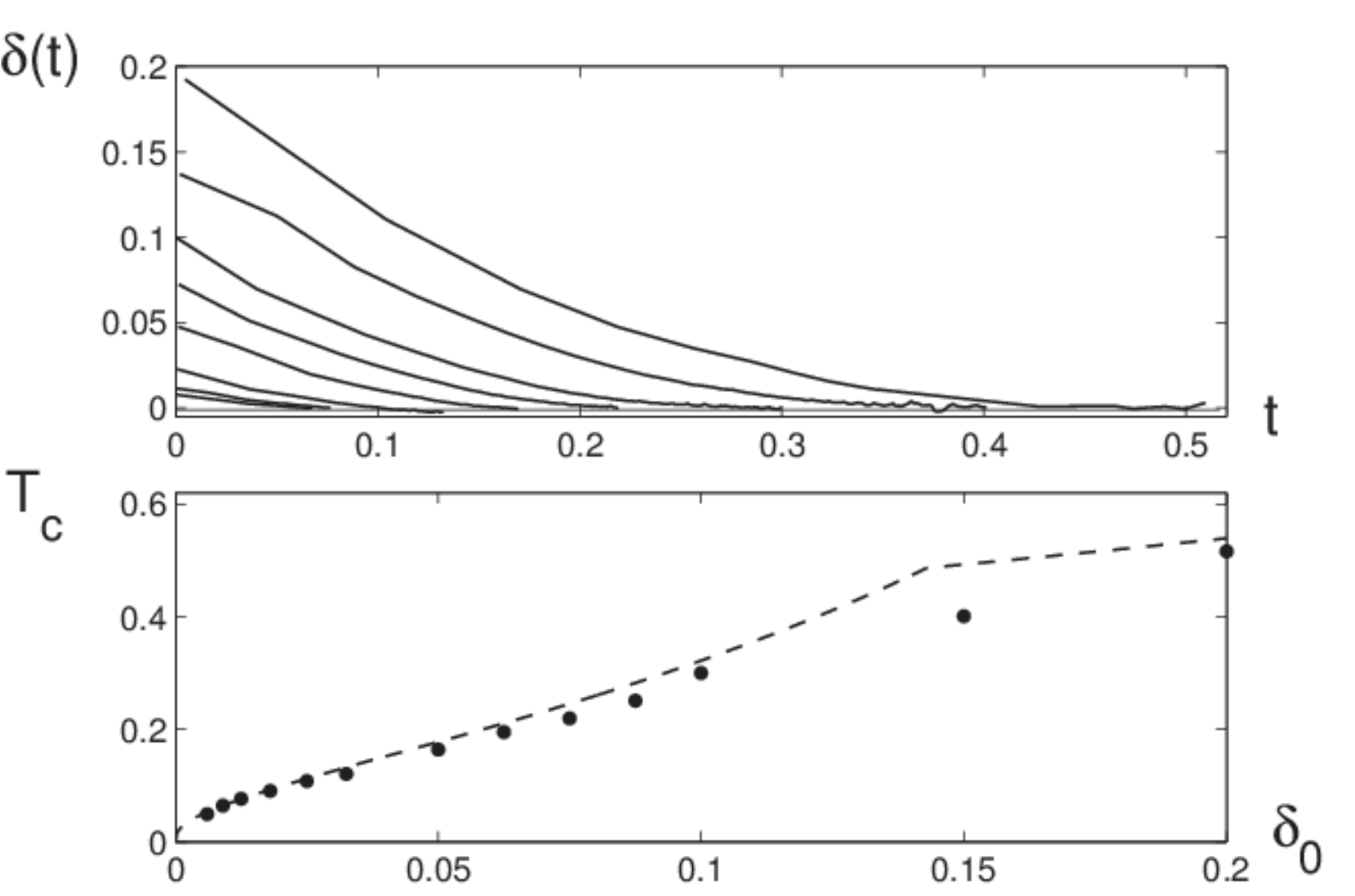}
\caption{{At the top the
distances from the real axis of the singularity for Prandtl's
equation at the location $Y=5$ with initial datum given by
\eqref{perturb}, for different initial $\delta_0$s.
At the bottom we compare the singularity times for different $\delta_0$
with the curve representing the $(\sup_{x,Y}|\pardx u_0|)^{-1}$.}
}\label{prandtl53alphadeltatimes}
\end{center}
\end{figure}

In Fig. \ref{prandtlSliding53} we use the sliding fitting
technique to show the shrinking of the analyticity strip and the
classification of the singularities at location $Y=5$ for various
$\delta_0=0.05$, $0.025$, and $0.0125$.
Note that, at the
singularity times, are all cubic--root singularities, therefore
stronger than the respective singularities for Burger's equation.
The results of the sliding fitting are in agreement with the results obtained
fitting the whole spectrum represented (only the $\delta(t)$) at the top of Fig.
\ref{prandtl53alphadeltatimes}.

At the bottom of Fig \ref{prandtl53alphadeltatimes}
we plot (with a dashed line) the curve  giving the dependence on $\delta_0$
of $1/\sup_{x,Y}|\pardx u_0|$, where $u_0$ are the initial data expressed by \eqref{iniperturb}.
The computed singularity times seem to lay quite close to this curve.
Therefore it seems to emerge a behavior of the singularity times similar
to the behavior of  Burger's equation.
Our findings reinforce the
idea that the generation of the separation singularity for
Prandtl's equations is analogous to the wave steepening for
compressible flow, and its connection to Burger's equation
(see \cite{VD81}).

Moreover we found a class of initial data, uniformly bounded in
norm $H^1$, for which our simulations seem to show that the singularity
time can be made arbitrarily short. At the singularity time the
solution has a strong discontinuity and it is not in $H^1$.

\begin{figure}[ht]
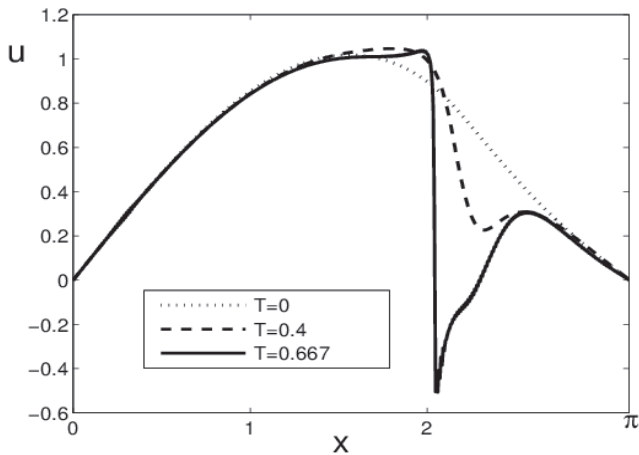

\leftline{\epsfxsize=6.5cm \epsfysize=5.1cm
\epsfbox{figure/upra52p05.eps}}\vskip-5.1cm\rightline{\epsfxsize=7.0cm
\epsfysize=5.0cm \epsfbox{figure/dastar52p05.eps}} \caption{The
solution of Prandtl's equation at the location $Y=5$ with initial
datum given by \eqref{prapert52} with $\delta_0=0.05$. To the left
one can follow the formation of a shock at $x\approx 2.14$ at the
location $Y=5$. To the right the results of the singularity
tracking. One can see that at $t\approx 0.667$ the solution
develops a cubic--root singularity at the real coordinate $x^\star
\approx 2.14$.}\label{pra52}
\end{figure}

\begin{figure}[ht]
\begin{center}
\includegraphics[height=6.5cm,width=9.5cm]{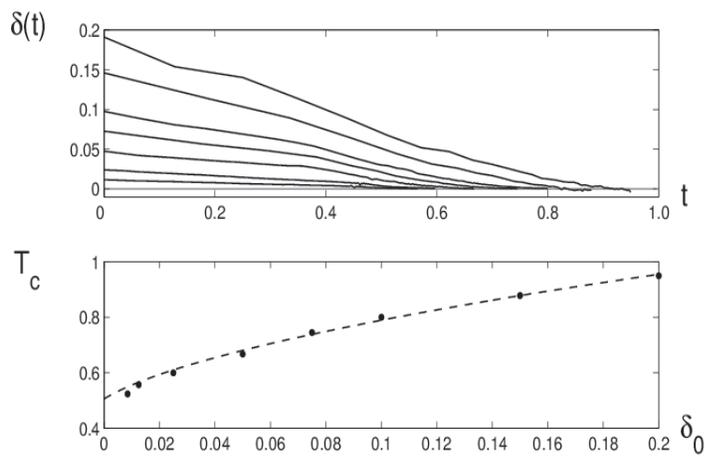} \caption{{At the top the
distances from the real axis of the singularity for Prandtl's
equation at the location $Y=5$ with initial datum given by
\eqref{prapert52}, for different initial $\delta_0$s.
At the bottom the singularity times for different $\delta_0$s.
The singularity time seem to lay on the curve $(\sup_{x,Y}|\pardx u_0|)^{-1}$.
This would mean that the singularity time remains bounded away from zero when
$\delta_0\rightarrow 0$.}}\label{pra52timedelta}
\end{center}
\end{figure}

If one imposes an initial datum with a weaker singularity one gets a
different behavior of the singularity times.
In fact, we impose  the  initial data \eqref{iniperturb}, where $U_{\delta_0}$
is now given by:
\begin{equation}\label{prapert52}
U_{\delta_0}(x)=\sum_{2\leq|k|\leq K/2}
-i\frac{k}{|k|}\frac{e^{-\delta_0 |k|}}{|k|^{5/2}}  \cos{(x^*
k)}e^{ikx}.
\end{equation}
These initial data are  uniformly bounded in $H^{3/2}$, for
each initial $\delta_0\geq 0$; they develop a cube--root singularity within
a finite time, as one can see in Fig. \ref{pra52}, where we report the
case of $\delta_0=0.05$.
In Fig. \ref{pra52timedelta} one can see that, also in this case, the singularity times
seem to be related to  $1/\sup_{x,Y}|\pardx u_0|$.
This would mean that, in this class of initial data, the blow--up time
of the derivative of the solution does not go to zero when $\delta_0\rightarrow 0$.

\section{The regularized Prandtl's equation}\label{regularized}
\setcounter{equation}{0}

We now consider the effects of a small viscosity in the streamwise direction on
the process of the singularity formation.
Namely we consider the equations:
\bea
\pardt u+u\pardx u +v\pardY u-U_{ }\pardx U_{ }&=&\pardYY u +\nu \pardxx u\qquad U=\sin{(x)}
\label{Pv1}\\
\pardx u +\pardY v &=&0  \label{Pv2}\\
u(x,0,t)=v(x,0,t)&=&0  \label{Pv3}\\
u(x,\infty,0)&=&U_{ }  \label{Pv4}
\eea
with initial datum:
\be
u(x,y,0)=\sin{(x)} \, . \label{Pv5}
\ee
We solve the above equation with the methods we have used for the
Prandtl's equation (described in Section \ref{NM}) plus the usual spectral
discretization of the viscous term $(\pardxx u)_k=-k^2u_k$ treated implicitly in time.
In Table \ref{convLW122viscous} we report the convergence properties
of the scheme based on the Lax--Wendroff approximation of the $Y$--derivative
and on the IMEX scheme $(1,2,2)$ of \eqref{122}.
The size of the time step  is adapted to have the CFL condition satisfied.
In Table \ref{convLW122viscous} we have chosen the initial datum $u_0=\sin(x)$.
One can see that at $T=3.5$, well after the VDS singularity time,
the scheme has good convergence properties, at least below a certain resolution.

\begin{figure}[ht]
\includegraphics[height=5.5cm,width=13.5cm]{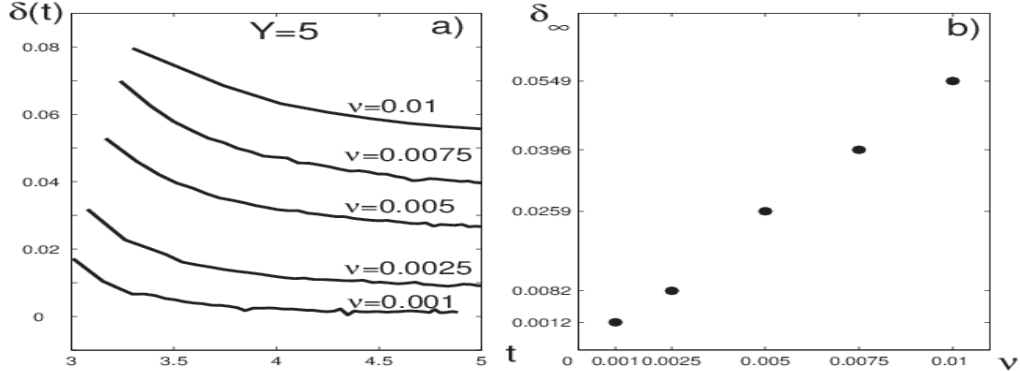} \caption{Imposing the
impulsively started disk datum, one can see that the singularities
do not hit the real axis (at least up to times $T=5$, well after
the VDS singularity time $t=3$) but stabilize at a distance
$\delta_\infty$ which depends on the amount of viscosity in the
streamwise direction. To the right it seems to emerge a linear
dependence of $\delta_\infty$ (which is related to the number of
Fourier modes effectively excited) on the viscosity $\nu$.
}\label{prandtl_deltavisctemp}
\end{figure}

In fact, if  we now follow the distance of the singularity from the real axis of the
singularity, we get the results of Fig. \ref{prandtl_deltavisctemp}, where the
results for different values of the viscosity are represented.
The resolution used for the computation of Fig.  \ref{prandtl_deltavisctemp}
is $K=1024$ in $x$ and $M=800$ in $Y$.
For the computations we have used a $32$-digits precision.

\begin{table}[ht]
\begin{scriptsize}
\begin{center}
\caption{Convergence of the numerical scheme based on the
Lax--Wendroff approximation of $\pardY G(u)$ and on the IMEX
scheme \eqref{122} for the solution of the regularized Prandtl's
equation. here the viscosity  is $\nu=0.01$. We show the convergence
properties at time $T=2$ and at time $T=3.5$, well after the VDS
singularity time. The initial datum is $u_0=\sin{x}$.}
\label{convLW122viscous} \vskip.1cm
\begin{tabular}{|c|c|c|c||c|c|c|}\hline
Grid   & \multicolumn{3}{|c||}{$T=2.0$} &\multicolumn{3}{|c|}{$T=3.5$} \\  \hline
($ K\times M$) & $L^2$    & $L^1$     & $L^\infty$
& $L^2$    & $L^1$     & $L^\infty$
\\ \hline
$ 32\times 127$         & $8.27\cdot 10^{-4}$ & $3.65\cdot 10^{-4}$   & 0.0063
                        & 0.081               & 0.045                 & 0.3409
\\ \hline
$ 64\times 255$  & $2.86\cdot 10^{-4}$   & $1.11\cdot 10^{-4}$ & $0.0020$
                 & $0.04$                & $0.022$             & $0.322$
 \\ \hline
$128\times 511$ & $1.17\cdot 10^{-4}$    & $4.63\cdot 10^{-5}$ & $8.2\cdot 10^{-4}$
                & 0.014                  & 0.0067              & 0.1691
\\ \hline
$256\times 1023$ & $5.24\cdot 10^{-5}$    & $2.21\cdot 10^{-5}$ & $3.75\cdot 10^{-4}$
                & 0.0022                  & $9.9\cdot 10^{-4}$  & 0.028

 \\ \hline
\end{tabular}
\end{center}
\end{scriptsize}
\end{table}

\begin{figure}[ht]
\begin{center}
\includegraphics[height=7.5cm,width=11.5cm]{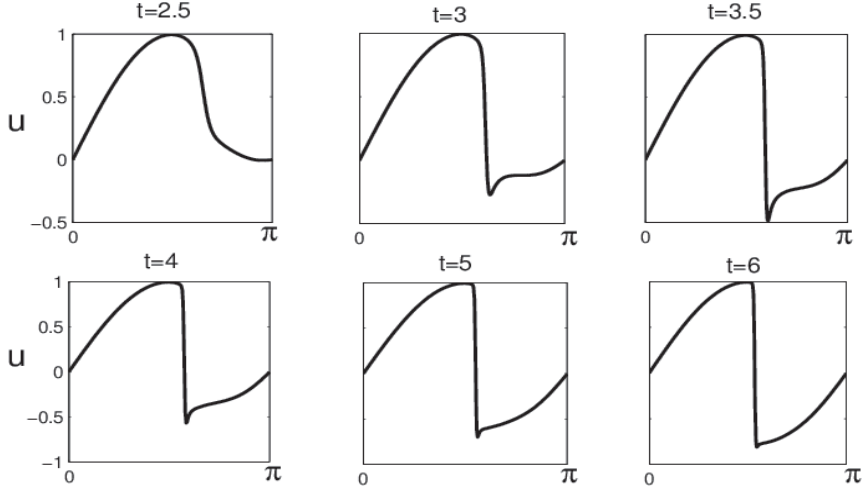} \caption{The solution of the
regularized Prandtl's equations starting from the initial datum
$u_0=\sin{(x)}$. The effect of the viscosity $\nu=0.005$ in the
streamwise direction is to regularize the solution. One can see
the steepening of the solution at $Y=5$ and the formation of a
regularized shock. }\label{Upravisc0p005}
\end{center}
\end{figure}

One can therefore see that the regularized Prandtl's equations, solved up to time $T=5$
well after the VDS singularity time, stay regular.
Moreover the singularity seems to stabilize at a distance of the real axis that
depends on the viscosity in $x$.
Plotting the distances that the singularities reach asymptotically in time
versus the viscosity it seems to emerge a linear dependence.
This suggests that, if $d$ is the number of degrees of freedom necessary to describe the
behavior of the regularized Prandtl's equations, then $d\sim \nu^{-1}$.
In Fig. \ref{Upravisc0p005} we plot the solution $u$ at the location $Y=5$
at different times. One can see how the solution steepens to form a regularized shock.

\begin{figure}[ht]
\includegraphics[height=9.5cm,width=13.5cm]{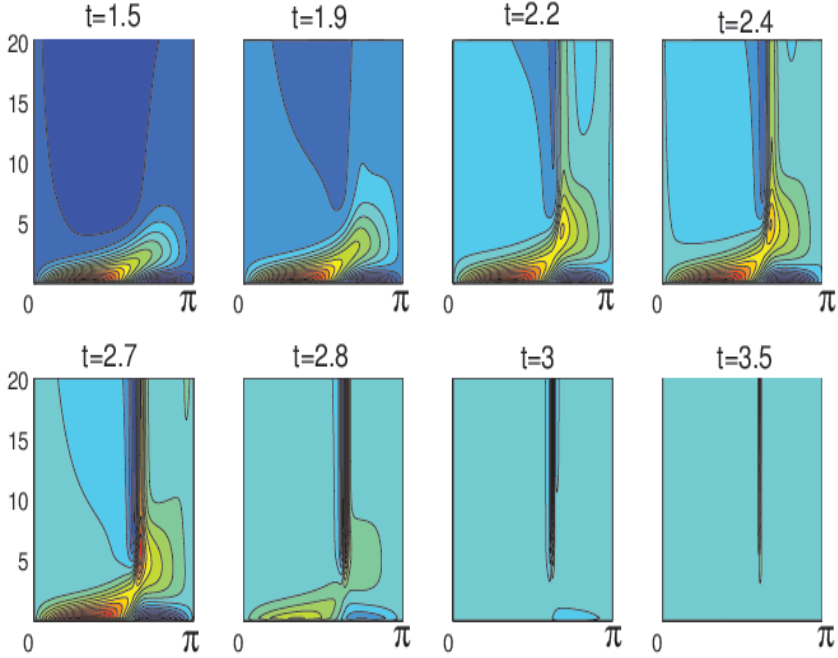}
\caption{Vorticity contours for the regularized Prandtl's equations with $\nu=0.005$
starting from the initial datum $u(x,Y,t=0)=\sin{(x)}$. The detachment of the vorticity
from the boundary is evident and the formation of two strips of high vorticity
of opposite sign is reminiscent of VDS singularity.
}\label{vortVDSvisc005}
\end{figure}

In Fig. \ref{vortVDSvisc005} we have represented the contours of
the vorticity $\omega =-\pardY u/\sqrt{\nu}+\sqrt{\nu}\pardx v$ at
various times. Vorticity detaches from the boundary and,
asymptotically, two strips of vorticity of opposite sign are
formed. These two strips are the result of the steepening of the
solution throughout the boundary layer as shown in Fig.
\ref{Upravisc0p005} at $Y=5$.

\begin{figure}[ht]
\includegraphics[height=9.5cm,width=13.5cm]{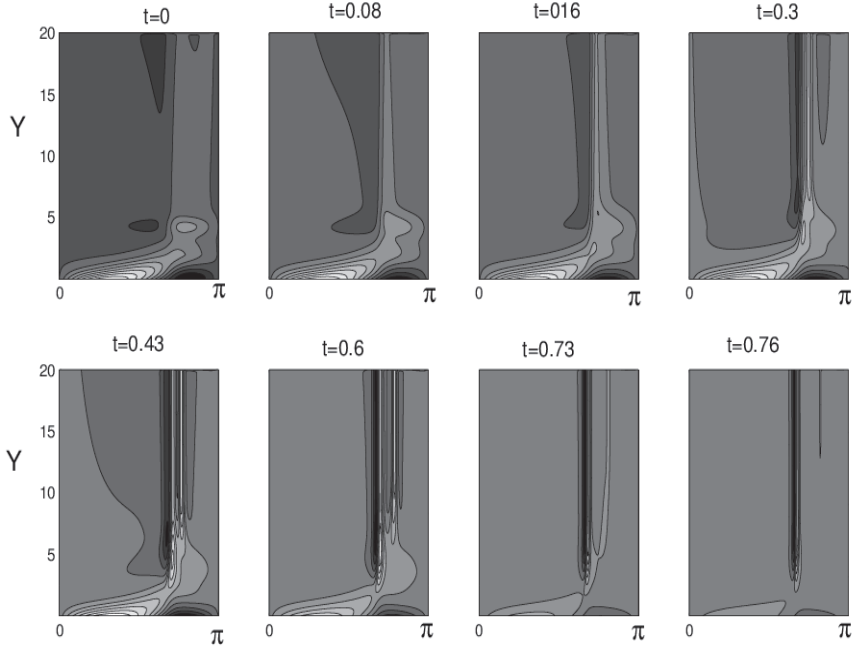} \caption{Instantaneous
vorticity $\omega=-\pardY u/\sqrt{\nu}+\sqrt{\nu}\pardx v$
contours of the solution of the regularized Prandtl's equations
for $\nu=0.005$ having imposed the initial datum
\eqref{iniperturb} with $\delta_0=0.2$. Looking at $t=0$ it
clarifies that the effect of the analytic perturbation
$U_{\delta_0}$ is to place a pair of counter-rotating vortices
inside the boundary layer. Notice how vorticity detaches from the
boundary to feed one of the two vortices. Asymptotically  it forms
two strips of vorticity of opposite sign.
}\label{vortv005deltap2alph53}
\end{figure}

In Fig. \ref{vortv005deltap2alph53} we see the vorticity contours starting from the
initial datum \eqref{iniperturb} with $\delta_0=0.2$.
Comparing the vorticity at $t=0$ in Fig. \ref{vortv005deltap2alph53} with the
vorticity at $t=1.5$ of Fig. \ref{vortVDSvisc005} one realizes that the effect
of the complex singularity we have added to the VDS datum at $t=1.5$ reveals
as a couple of counter rotating vortices places in the boundary layer.
These two vortices speed up the process of the vorticity detachment from
the boundary and the process of the formation of the high vorticity strip,
as it is apparent from a comparison between Fig. \ref{vortVDSvisc005}
and Fig. \ref{vortv005deltap2alph53}.

We finally notice that to resolve the regularized Prandtl's
equation it is necessary to use a number $K$ of modes in $x$ such that
$K\gtrsim 1/\delta_\infty$.
In fact if the singularity gets to a distance from the real axis smaller than
the resolution in $x$, this would result in the formation of spurious oscillations
and in the ultimate  blow up of the computed solution.

\section{Conclusions}\label{Con}
\setcounter{equation}{0} The singularity tracking method has
revealed an effective tool to investigate the dynamics of the
boundary layer equations. The process leading to  Van Dommelen
and Shen's singularity has revealed itself as the results of a
complex cubic--root singularity (coming from infinity) hitting the
real axis (of the complexified $x$ variable) at time $t\approx 3$.
We have seen how this process can be accelerated perturbing the
boundary layer solution with a datum containing a singularity (the
dipole singularity) at distance $\delta_0$ from the real axis.
Moreover we have seen that it seems possible to make the
singularity time short at will picking the initial datum in a
class uniformly bounded in the norm $H^1_x\times C^2_Y$. The
critical time for Prandtl's equations (in its dependence from
$\delta_0$ as $\delta_0$ is taken smaller) seems to behave as it
does the critical time of Burger's equation for similar
initial data, see Fig. \ref{prandtl53alphadeltatimes} at the
bottom. We stress that at the singularity time the solution is not
in $H^1$, being the singularity of the cubic--root type, see Fig.s
\ref{prandtlSliding53} at the right. The well posedness of the
Prandtl's equations has been proved for analytic (in $x$) initial
data or for monotonous initial data. The question of the well
posedness (or of the ill posedness) in Sobolev spaces is an
important question that remains open. The findings of this paper
(especially if supported by calculations at higher resolution
which would allow to consider initial singularities increasingly
closer to the real axis) suggest that the well posedness (if
possible) needs regularity in $x$ higher than $H^1$.

The introduction in Prandtl's equations of a small viscosity in the streamwise
direction prevents the formation of the singularity (at least for the viscosities
that we have been able to test).
In fact the complex singularity instead of hitting the real axis, after a transient
during which it gets closer to the real axis, it stabilizes at a distance $\delta_\infty$
which depends (linearly, it seems) on the amount of regularizing viscosity.
Being $\delta_\infty$ related to the number of the Fourier modes effectively excited,
this indicates a finite dimensional dynamics of the 2D regularized Prandtl's equations.
We believe it would be interesting to confirm (or disprove) this at smaller viscosity
(which requires higher resolution) and/or in a 3D setting.
We finally notice that the solution of the regularized Prandtl's equations
for relatively large times requires, due to the growth of the boundary layer,
to consider a larger computational domain in the normal direction; or, alternatively,
the coupling with the Euler or the Navier--Stokes equations at the
outer boundary.
This, and other topics, will be the subject of future work.

\section*{Acknowledgments} The authors would like to thank R.Caflisch for several
enlightening discussions on the topics of this paper.
We also thank D.Bailey and H.Yozo for the help in the use of the ARPREC package.
This work has been partially supported by the INDAM and by the PRIN grant
``Nonlinear propagation and stability in thermodynamical processes of continuous media".

\appendix
\section{Appendix: E and Engquist's singularity}

In this appendix we analize the singularity formation for Prandtl's equation for an analytic initial datum given firstly by E and Engquist \cite{EE97},
and to determinate its complex characterization.
We remember that the work of E and Engquist (EE) is the only analytic existing result about the blow--up for solutions of Prandtl's equation at a finite time. 
We point out some differences between EE singularity and the separation singularity which accous in VanDommelen and Shen situation analized in section 3. We consider here a periodic version of the work of E and Engquist, which is simple to obtain.

We consider the periodic Prandtl's equation in the domain $D=[-\pi, \pi]\times[0,\infty]$
\bea
\pardt u +u\pardx u+v\pardY u &=&\pardYY u ,\label{P_absnt.1} \\
\pardx u+\pardY v &=& 0 ,\label{P_absnt.2} \\
u|_{Y=0} = v|_{Y=0} &=& 0, \qquad u\rightarrow 0 \quad \mbox{when} \quad Y\rightarrow\infty\;  \label{P_absnt.3} \\
u|_{x=-\pi} &=& u|_{x=\pi}, \label{P_absnt.4} \\
u|_{t=0} &=& u_0 = -\sin(x) b_0(x,Y), \label{init_absnt} 
\eea
where the outer Euler inviscid flow is imposed to be zero, and where
$b_0 \geq 0$ is a regular function of $x$ and $y$ in $D$, not necessarily periodic in $x$, which satisfies the following condition
\be
E(a_0)=\int_{0}^{\infty} \left( \frac{1}{2} (\pardY a_0)^2 -\frac{1}{4} a_0^3 \right) dY < 0; \label{EE_cond}
\ee
with $a_0=b|_{x=0}\geq 0$. E and Engquist  prove that there exists a finite time $T$ such that the $x$--derivative of the solution $u$ of Prandtl equations blow up.
In particular, this means that we have $\lim_{t\rightarrow T}\sup_{Y>0} |\frac{u(x,Y,t)}{\sin x}|_{x=0}=+\infty$ or otherwise
\be
\lim_{t\rightarrow T} \sup_{Y>0} |\pardx u(0,Y,t)|=+\infty .\label{EE_blowup}
\ee

\begin{figure}[ht]
\begin{center}
\includegraphics[height=8.5cm,width=15.5cm]{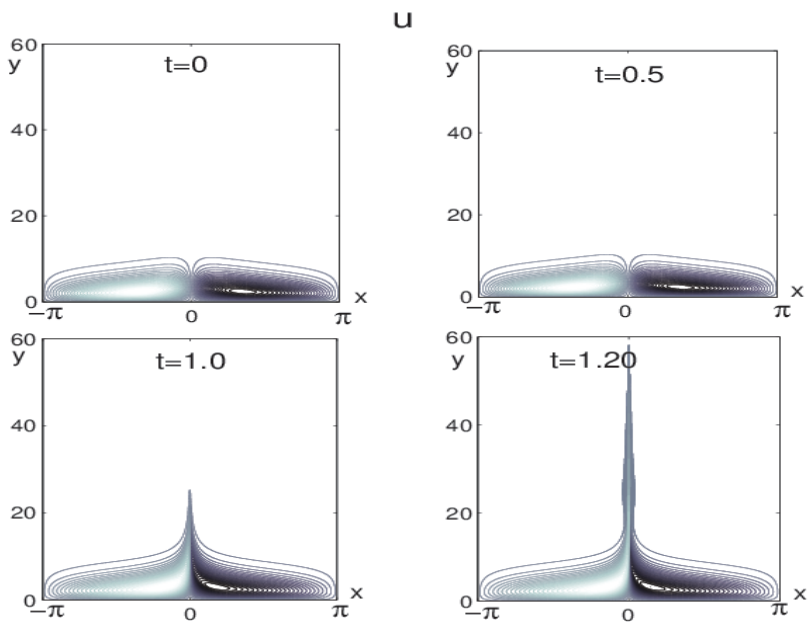} \caption{ The streamwise velocity $u$ of
Prandtl's equation with initial datum given by \eqref{a_init_num}. One can follow the formation of a shock at $t\approx 1.20$.}\label{EE_fig_u}
\end{center}
\end{figure}

Like in section 3 for the VDS's initial datum, we solve numerically Prandtl's equation \eqref{P_absnt.1}--\eqref{P_absnt.4}, with inital datum given by \eqref{init_absnt} and
\be
b_0(x,Y)=a_0(Y)=\frac{1}{4} Y^2 e^{-(Y-2)}, \label{a_init_num}
\ee
which satisfies condition \eqref{EE_cond}. In figures \eqref{EE_fig_u} it is show the behavior in time of the solutions for Prandtl's equation with initial datum given by \eqref{init_absnt} and \eqref{a_init_num}. In this case the flows corresponds to flows impinging from the left and right at the line $x=0$, which is the rear stagnation point.

At time $t\approx 1.20$ Gibb's phenomenons are visible for the velocity $u$, so we argue that Prandtl's equation develops a singularity at this time. As the singularity forms, the singular structure seems to be convected to $Y=+\infty$, making the numerical computations difficult. We have considered the computational domain as $[-\pi,\pi]\times[0,Y_{\mbox{\tiny MAX}}]$, and we chose $Y_{\mbox{\tiny MAX}}=60$ where the velocity $u$ is of the order of $10^{-18}$ at time $t\approx 1.20$.

\begin{figure}[ht]
\begin{center}
\includegraphics[height=7.5cm,width=13.5cm]{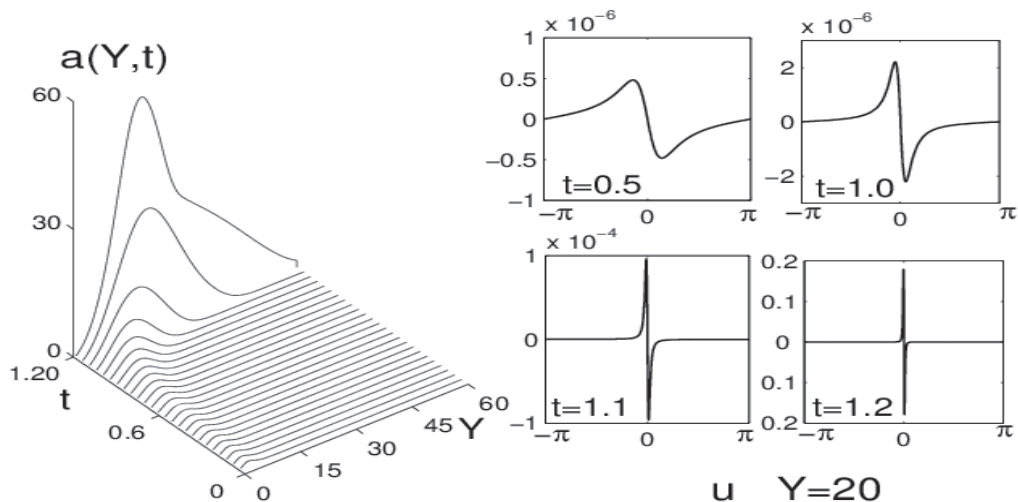} \caption{ On the right figure the behavior of the function $a(Y,t)=\pardx u(0,Y,t)$ for solution of Prandtl's equation with initial datum \eqref{a_init_num}. On the left the profiles of $u$ at the cut $Y=20$ and at different times.} \label{EE_fig_uux}
\end{center}
\end{figure}

In the figure on the left of Fig.\ref{EE_fig_uux} we show the behavior of $a(Y,t)=\pardx u(0,Y,t)$, as one can see at time $t\approx 1.20$ the maximum grows around the location $Y\approx20$. On the right of figure Fig.\ref{EE_fig_uux} we show the profile at different times of $u$ at the location $Y=20$ where the maximum of $a$ seems to be reach.

\begin{figure}[ht]
\begin{center}
\includegraphics[height=5.0cm,width=11.5cm]{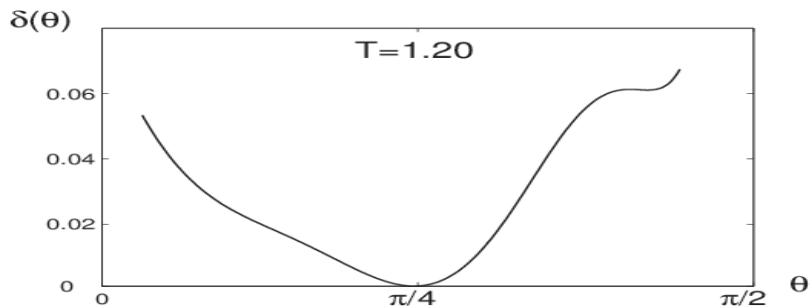} \caption{The profile of the distance $\delta$ at the singularity time w.r.t. the angle $\theta$ for the solution of
Prandtl's equation with initial datum \eqref{a_init_num}, at the singularity time $t=1.20$. The distance $\delta$ seems to reach its minimum at $\approx \pi/4$.} \label{EE_fig_deltaTheta}
\end{center}
\end{figure}

\begin{figure}[ht]
\begin{center}
\includegraphics[height=8.0cm,width=12.5cm]{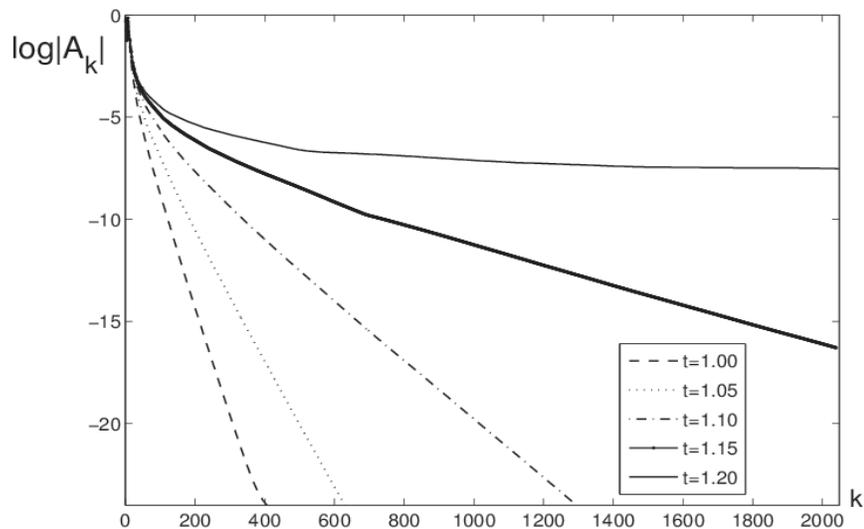} 
\caption{
The time evolution of the shell--summed amplitude of the fourier spectrum for Prandtl's equation with initial datum given by \eqref{a_init_num}.} \label{EE_fig_shellSum}
\end{center}
\end{figure}
\begin{figure}[ht]
\begin{center}
\includegraphics[height=8.5cm,width=11.5cm]{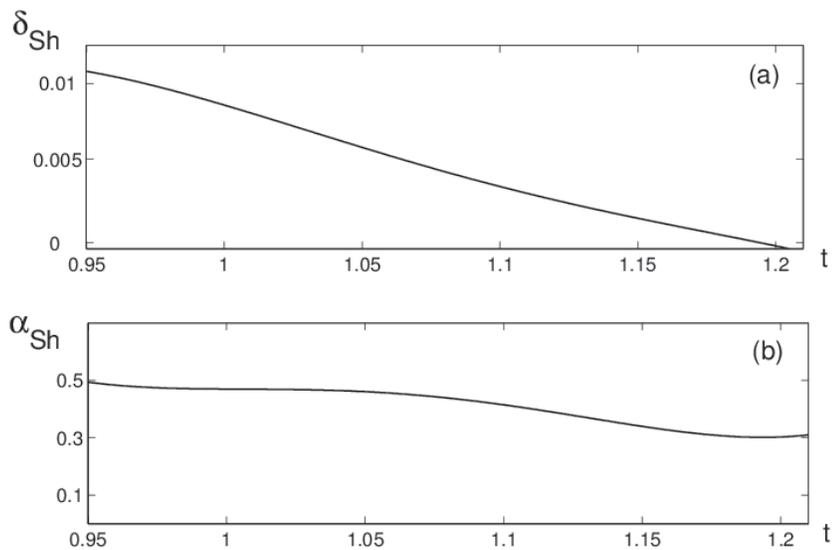} \caption{The singularity tracking results for the solution of
Prandtl's equation with initial datum given by \eqref{a_init_num}. In $(a)$ and $(b)$ the respective evolutions in time of the distance $\delta$ and the algebraic factor $\alpha$.} \label{EE_fig_shellSum_Fitt}
\end{center}
\end{figure}

In figure Fig.\ref{EE_fig_deltaTheta} it is shown the behavior of the strip of analyticity $\delta$ with respect to the angle $\theta$ at the singularity time $t=1.20$. As one can see, this case is quite different from the VDS--singularity. In this case the distance $\delta$ seems to reach a minimum at $\approx \pi/4$, and this means that the influence for the complex singularity by the $Y$ variable is not neglecting for the EE singularity.
In this case the technique of singularity tracking to detecting the location and the algebraic type of the complex singularity at different cuts of $Y$ does not work, then we analyze the behavior in time of the corresponding shell--summed Fourier amplitude \cite{MBF05} $A_K$ defined in the subsection 2.4. In Fig.\ref{EE_fig_shellSum} we show the evolution of $A_K$ at different times.
In Fig.\ref{EE_fig_shellSum_Fitt}, one can see the distance $\delta$ which reach the real axis at the time $t=1.20$ and $\alpha$ is equal to $1/3$. This confirm the formation of a shock singularity in the $x$ derivative for the solution of Prandtl's equation with an EE type initial datum given by \eqref{a_init_num}.

Finally, we point out that if we choose another type of exponential decaying at infinity, the results are similar.
As a conclusion we said that the EE singularity developed by the initial datum \eqref{a_init_num}, analyzed in this section seems to be a shock--singularity in the $\pardx u$, like in the classical VDS case. Moreover, the situation is different from the VDS case for some aspects. First of all, no recirculation region forms. The flow remains separated at the rear stagnation point where the singularity develops, different from the VDS case where the flow separates from the boundary at approximately $3 \pi /4$ from the front stagnation point. As second step, the complex singularity in the EE case depends not only by the streamwise variable like in VDS, but also by the normal variable, as we point out in this appendix.


\section{Appendix: Perturbation at the Euler outer flow}

In this appendix we shall investigate the response to perturbation at both the initial datum and Euler outer flow of the VDS case, adding an analytic function in $x$ that has a singularity at distance $\delta_0$ from the real axis, as doing in section 4.
Also in this situation, the addition of a perturbation accelerates
the formation of the singularity. Moreover, the algebraic structure of shock singularity remains stable, i.e. differents initial configurations devolops singularity at different times but all have the same shock type of singularity observed for the VDS case.

We consider the following family of initial data and the Euler outer flows $U$ for Prandtl's
equations \eqref{P1}--\eqref{P4}, with 
\be 
u_0(x,Y)=U(x)=U_{(\delta_0,\alpha_{in},x^{*},\sigma)}\; , \label{iniperturb_app} 
\ee 
where
\begin{equation}\label{perturb}
U_{(\delta_0,\alpha_{in},x^{*},\sigma)}(x)=\sin(x) + \sigma \sum_{2\leq|k|\leq K/2}
-i\frac{k}{|k|}\frac{e^{-\delta_0 |k|}}{|k|^{\alpha_{in}+1}}  \cos{(x^*
k)}e^{ikx}.
\end{equation}

We now solve Prandtl's equations \eqref{P1}--\eqref{P4} with initial datum and Euler outer flow given by
\eqref{iniperturb_app} \eqref{perturb} with differentes $\delta_0$, $\alpha_{in}$, $\sigma$ and $x^{*}$.
The calculations we shall present here have been performed using a fully
spectral method with resolution $2048^2$.

\begin{figure}[]
\begin{center}
\includegraphics[height=8.5cm,width=9.cm]{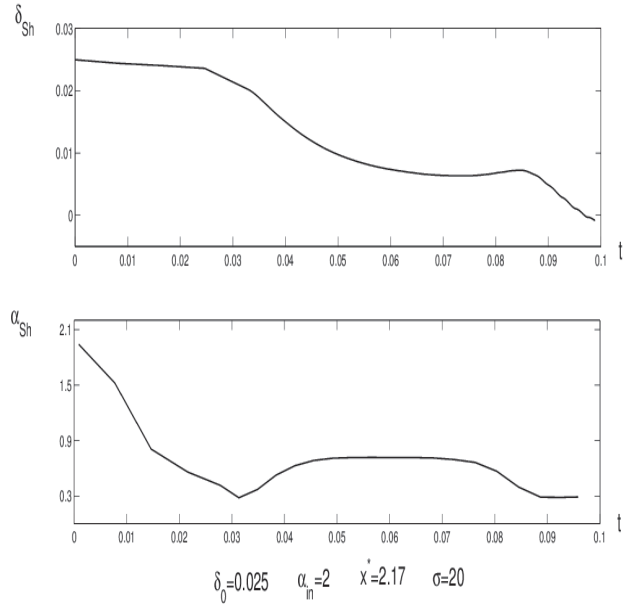}
\caption{The behavior in time of the width of the analyticity strip for initial and Euler outer flow given by \eqref{iniperturb_app} with $\delta_0=0.025$, $\alpha_{in}=2$, $\sigma=20$ and $x^{*}=2.17$.
The singularity time is $t_c\approx 0.098$, and the singularity is of cubic-root
type.
}\label{shell_fitting_delta_alpha_pertinf025}
\end{center}
\end{figure}

\begin{figure}[]
\begin{center}
\includegraphics[height=8.5cm,width=9.cm]{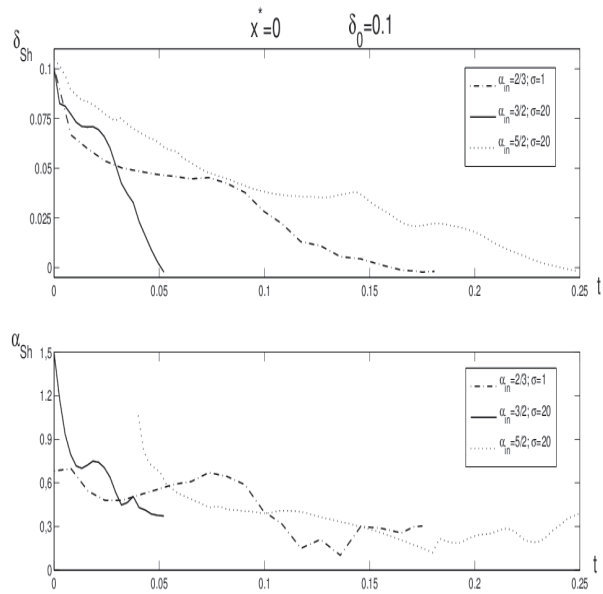}
\caption{The behavior in time of the width of the analyticity strip for initial and Euler outer flow given by \eqref{iniperturb_app} with differents initial configurations of $\alpha_{in}$ and $\sigma$, at the same $x^{*}=0$ and $\delta_0 =0.1$. All the singularities are of cubic-root type.
}\label{shell_fitting_delta_alpha_pertinf_xstar0}
\end{center}
\end{figure}

We present in Fig. \ref{shell_fitting_delta_alpha_pertinf025} and Fig. \ref{shell_fitting_delta_alpha_pertinf_xstar0}, the fitting of the shell--summed amplitude for differents initial configurations.
As one can see, all the solutions lose analyticity at finite time and have
the same cubic-root singularity.

\begin{figure}[]
\begin{center}
\includegraphics[height=7.5cm,width=9.5cm]{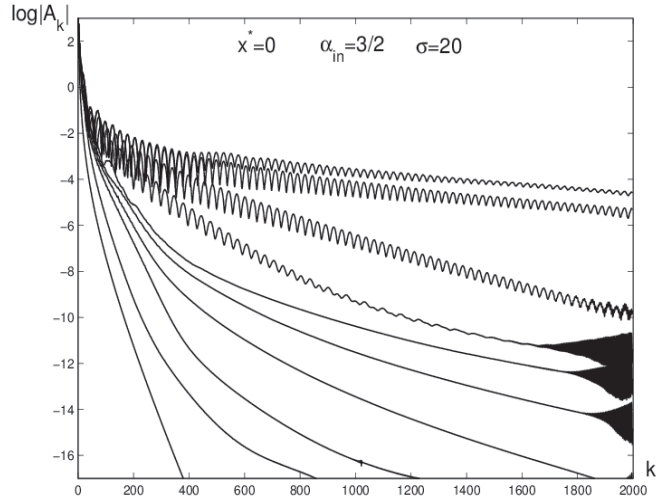}
\caption{  {The behavior in time of the shell summed amplitude up
to the singularity time, for initial and Euler outer flow given by \eqref{iniperturb_app} with $\delta_0=0.1$, $\alpha_{in}=3/2$, $\sigma=20$ and $x^{*}=0$.
}}\label{shell_pertinf}
\end{center}
\end{figure}

We point out that for initial $x^{*}=0$, the $x$--location of the singularity is different from zero, as one can see in Fig. \ref{shell_pertinf} were we show the shell-summed amplitudes with $\delta_0=0.1$, $\alpha_{in}=3/2$, $\sigma=20$ and $x^{*}=0$. It is evident
the loss of exponential decay and the formation of oscillations at the singularity time.

\bibliographystyle{siam}

\bibliography{prandtl}

\begin{thebibliography}{10}

\bibitem{AS72}
{\sc M.~Abramowitz and I.~A. Stegun}, {\em {Handbook of Mathematical
  Functions}}, Dover, New York, 1972.

\bibitem{Asano88b}
{\sc K.~Asano}, {\em {Zero Viscosity Limit of the Incompressible Navier--Stokes
  Equations 1 and 2}}, preprint,  (1988).

\bibitem{ARS97}
{\sc U.~Ascher, S.~Ruuth, and R.~Spiteri}, {\em Implicit--explicit
  {R}unge--{K}utta methods for time--dependent partial differential equations},
  Appl. Numer. Math., 25 (1997), pp.~151--167.

\bibitem{ARPREC}
{\sc D.~Bailey, H.~Yozo, X.~Li, and B.~Thompson}.
\newblock ARPREC: An arbitrary precision computation package, (September 1,
  2002). Lawrence Berkeley National Laboratory. Paper LBNL-53651.

\bibitem{BT07}
{\sc C.~Bardos and {\`E}.~S. Titi}, {\em Euler equations for an ideal
  incompressible fluid}, Uspekhi Mat. Nauk, 62 (2007), pp.~5--46.

\bibitem{Boyd01}
{\sc J.~P. Boyd}, {\em Chebyshev and {F}ourier spectral methods}, Dover
  Publications Inc., Mineola, NY, second~ed., 2001.

\bibitem{BW01}
{\sc K.~Brinckman and J.~Walker}, {\em Instability in a viscous flow driven by
  streamwise vortices}, J. Fluid. Mech., 432 (2001), pp.~127--166.

\bibitem{Caf93}
{\sc R.~Caflisch}, {\em Singularity formation for complex solutions of the {3D}
  incompressible euler equations}, Phisica D, 67 (1993), pp.~1--18.

\bibitem{CS00}
{\sc R.~Caflisch and M.~Sammartino}, {\em {Existence} and {Singularity } for
  the {Prandtl} {Boundary} {Layer} {Equations}}, Z. Angew Math. Mech, 80
  (2000), pp.~733--744.

\bibitem{CLS01}
{\sc M.~Cannone, M.~Lombardo, and M.~Sammartino}, {\em {Existence} and
  {Uniqueness} for the {Prandtl} {Equations}}, C.R. Acad. Sci. Paris S\'er I
  Math, 332 (2001), pp.~277--282.

\bibitem{CHQZ06}
{\sc C.~Canuto, M.~Y. Hussaini, A.~Quarteroni, and T.~A. Zang}, {\em Spectral
  methods: Fundamentals in single domains}, Scientific Computation,
  Springer-Verlag, Berlin, 2006.

\bibitem{CKP66}
{\sc G.~Carrier, M.~Krook, and C.~Pearson}, {\em {Functions of a Complex
  Variable: Theory and Technique}}, McGraw--Hill, New York, 1966.

\bibitem{Cas00}
{\sc K.~Cassel}, {\em A comparison of {N}avier-{S}tokes solutions with the
  theoretical description of unsteady separation}, R. Soc. Lond. Philos. Trans.
  Ser. A Math. Phys. Eng. Sci., 358 (2000), pp.~3207--3227.

\bibitem{Cow01}
{\sc S.~Cowley}, {\em {Laminar Boundary-Layer Theory: A 20th Century
  Paradox?}}, in Mechanics for a New Millennium, Proceedings of 20th ICTAM,
  H.~Aref and J.~Phillips, eds., New York, 2001, Springer, pp.~389--411.

\bibitem{CBT99}
{\sc S.~Cowley, G.~Baker, and S.~Tanveer}, {\em On the formation of {Moore}
  curvature singularities in vortex sheets}, J. Fluid. Mech., 378 (1999),
  pp.~233--267.

\bibitem{DLSS06}
{\sc G.~Della~Rocca, M.~C. Lombardo, M.~Sammartino, and V.~Sciacca}, {\em
  Singularity tracking for {C}amassa-{H}olm and {P}randtl's equations}, Appl.
  Numer. Math., 56 (2006), pp.~1108--1122.

\bibitem{E00}
{\sc W.~E}, {\em Boundary layer theory and the zero--viscosity limit of the
  {Navier--Stokes} equations}, Acta Math. Sin., 16 (2000), pp.~207--218.

\bibitem{EE97}
{\sc W.~E and B.~Engquist}, {\em Blowup of the {Solutions} to the {Unsteady
  Prandtl's Equations}}, Comm. Pure Appl. Math., 50 (1997), pp.~1287--1293.

\bibitem{FMB03}
{\sc U.~Frisch, T.~Matsumoto, and J.~Bec}, {\em {Singularities of Euler Flow?
  Not out of the Blue!}}, J. Stat. Phys., 113 (2003), pp.~761--781.

\bibitem{GPS98}
{\sc R.~Goldstein, A.~Pesci, and M.~Shelley}, {\em Instabilities and
  singularities in {Hele--Shaw} flow}, Physics of Fluids, 10 (1998),
  pp.~2701--2723.

\bibitem{Gren00a}
{\sc E.~Grenier}, {\em On the stability of boundary layers of incompressible
  euler equations}, J. Differential Equations, 164 (2000), pp.~180--222.

\bibitem{Hong02}
{\sc L.~Hong}, {\em A Numerical and Analytic Study of the Prandtl Equations},
  {PhD} dissertation, Univerity of California, Davis, Dept. of Mathematics,
  2002.

\bibitem{HoHu03}
{\sc L.~Hong and J.~Hunter}, {\em {Singularity Formation} and {Instability} in
  the {Unsteady Inviscid} and {Viscous Prandtl Equations}}, Comm. Math. Sci., 1
  (2003), pp.~293--316.

\bibitem{Ing84}
{\sc D.~Ingham}, {\em Unsteady separation}, J. Comp. Phys., 53 (1984),
  pp.~90--99.

\bibitem{Ka84}
{\sc T.~Kato}, {\em Remarks on the zero viscosity limit for nonstationary
  navier-stokes flows with boundary}, in Seminar on nonlinear partial
  differential equations, no.~2 in Math. Sci. Res. Inst. Publ., Springer, New
  York, 1984, pp.~85--98.

\bibitem{LCS03}
{\sc M.~Lombardo, M.~Cannone, and M.~Sammartino}, {\em {Well}--{Posedness} of
  the {Boundary} {Layer} {Equations}}, SIAM J. Math. Anal., 35 (2003),
  pp.~987--1004.

\bibitem{MBF05}
{\sc T.~Matsumoto, J.~Bec, and U.~Frisch}, {\em {The Analytic Structure of 2{D}
  Euler Flow at Short Times}}, Fluid Dyn. Res., 36 (2005), pp.~221--237.

\bibitem{OC05}
{\sc A.~Obabko and K.~Cassel}, {\em On the ejection--induced instability in
  {N}avier--{S}tokes solutions of unsteady separation}, Phyl. Trans. R. Soc. A,
  363 (2005), pp.~1189--1198.

\bibitem{O66}
{\sc O.~Oleinik}, {\em {On the Mathematical Theory of Boundary Layer for an
  Unsteady flow of Incompressible Fluid}}, J. Appl. Math. Mech., 30 (1966),
  pp.~951--974.

\bibitem{OS99}
{\sc O.~Oleinik and V.~Samokhin}, {\em Mathematical models in boundary layer
  theory}, vol.~15 of Applied Mathematics and Mathematical Computation, Chapman
  \& Hall/CRC, Boca Raton, FL, 1999.

\bibitem{PF07}
{\sc W.~Pauls and U.~Frisch}, {\em A {B}orel transform method for locating
  singularities of {T}aylor and {F}ourier series}, J. Stat. Phys., 127 (2007),
  pp.~1095--1119.

\bibitem{PMFB06}
{\sc W.~Pauls, T.~Matsumoto, U.~Frisch, and J.~Bec}, {\em {Nature of Complex
  Singularities for the 2D Euler Equation}}, Physica D, 219 (2006), pp.~40--59.

\bibitem{PS98}
{\sc M.~Pugh and M.~Shelley}, {\em { Singularity Formation in Thin Jets with
  Surface Tension}}, Comm. Pure Appl. Math., 51 (1998), pp.~733--795.

\bibitem{SC98a}
{\sc M.~Sammartino and R.~Caflisch}, {\em {Zero} {Viscosity} {Limit} for
  {Analytic} {Solutions} of the {Navier}--{Stokes} {Equations} on a
  {Half--Space} {I}: {Existence} for {Euler} and {Prandtl} {Equations}}, Comm.
  Math. Phys., 192 (1998), pp.~433--461.

\bibitem{SC98b}
\leavevmode\vrule height 2pt depth -1.6pt width 23pt, {\em Zero viscosity limit
  for analytic solutions of the {Navier--Stokes} equations on a half--space
  {II}: Construction of the {Navier}--{Stokes} solution}, Comm. Math. Phys.,
  192 (1998), pp.~463--491.

\bibitem{Sh92}
{\sc M.~Shelley}, {\em {A study of singularity formation in vortex--sheet
  motion by a spectrally accurate vortex method}}, J. Fluid. Mech., 244 (1992),
  pp.~493--526.

\bibitem{SSF83}
{\sc C.~Sulem, P.-L. Sulem, and H.~Frisch}, {\em {Tracing Complex Singularities
  with Spectral Methods}}, J. Comput. Phys., 50 (1983), p.~138–161.

\bibitem{TW97}
{\sc R.~Temam and X.~Wang}, {\em The convergence of the solutions of the
  {Navier-Stokes} equations to that of the {Euler} equations}, App. Math.
  Lett., 10 (1997), pp.~29--33.

\bibitem{VD81}
{\sc L.~{Van Dommelen}}, {\em Unsteady Boundary Layer Separation}, {PhD}
  dissertation, Cornell University, 1981.

\bibitem{VDC89}
{\sc L.~{Van Dommelen} and S.~Cowley}, {\em On the {L}agrangian description of
  unsteady boundary-layer separation. {I}.\ {G}eneral theory}, J. Fluid Mech.,
  210 (1990), pp.~593--626.

\bibitem{VDS80}
{\sc L.~{Van Dommelen} and S.~Shen}, {\em The {Spontaneous Generation} of the
  {Singularity} in a {Separating Laminar Boundary Layer}}, J. Comp. Phys., 38
  (1980), pp.~125--140.

\bibitem{VDS82}
\leavevmode\vrule height 2pt depth -1.6pt width 23pt, {\em {The Genesis of
  Separation}}, in Symposium on Numerical and Physical Aspects of Aerodynamic
  Flows, T.~Cebeci, ed., Springer, 1982, pp.~293--311.

\bibitem{XZ04}
{\sc Z.~Xin and L.~Zhang}, {\em {On the Global Existence of Solutions to the
  {Prandtl's} System}}, Adv. Math., 181 (2004), pp.~88--133.

\end{thebibliography}

\end{document}